\documentclass[11pt, twoside, nohyper]{JHEP}
\makeatletter
\newdimen\captionmargin
\newdimen\captionindent
\newdimen\captionwidth
\newcommand\captionfont{}
\newcommand\@captionlabel[1]{#1:\space}
\long\def\@makecaption#1#2{%
  \vskip\abovecaptionskip
  \captionwidth\hsize
  \advance\captionwidth -2\captionmargin
  \sbox\@tempboxa{\@captionlabel{#1}\captionfont #2}%
  \ifdim \wd\@tempboxa >\captionwidth
    \ifdim\captionindent>\z@
      \advance\captionwidth -\captionindent
      \hskip\captionindent
    \fi
    \hskip\captionmargin
    \parbox[t]{\captionwidth}{\leavevmode\hskip-\captionindent
      \@captionlabel{#1}\captionfont #2}%
  \else  
    \global \@minipagefalse
    \hb@xt@\hsize{\hfil\box\@tempboxa\hfil}%
  \fi
  \vskip\belowcaptionskip}
\renewcommand\thetable{\@arabic\c@table}
\makeatother

\newcommand{\be}[3]{\begin{equation}  \label{#1#2#3}}     
\newcommand{\ee}{ \end{equation}}
\newcommand{\ba}{\begin{array}}
\newcommand{\ea}{\end{array}}
\newcommand{\bea}{\begin{eqnarray}}
\newcommand{\eea}{\end{eqnarray}}
\newcommand{\NP}[3]{{\em Nucl. Phys.}{ \bf B#1#2#3}}

\newcommand{\eqn}[1]{(\ref{#1})}

\def\ve{\varepsilon}
\newcommand{\ft}[2]{{\textstyle\frac{#1}{#2}}}

\def\beq{\begin{equation}}
\def\eeq{\end{equation}}
\def\beqa{\begin{eqnarray}}
\def\eeqa{\end{eqnarray}}

\renewcommand{\a}{\alpha}
\renewcommand{\b}{\beta}

\renewcommand{\d}{\delta}
\newcommand{\pa}{\partial}
\newcommand{\g}{\gamma}
\newcommand{\G}{\Gamma}

\newcommand{\e}{\epsilon}

\renewcommand{\l}{\lambda}
\renewcommand{\L}{\Lambda}

\newcommand{\m}{\mu}

\newcommand{\n}{\nu}

\newcommand{\s}{\sigma}

\renewcommand{\o}{\omega}
\renewcommand{\O}{\Omega}

\title{Stationary BPS Solutions in $N=2$ Supergravity with $R^2$-Interactions}

\author{Gabriel Lopes Cardoso\\Spinoza Institute, Utrecht University,
  Utrecht, The Netherlands, \email{cardoso@phys.uu.nl}}
\author{Bernard de Wit\\Institute for Theoretical Physics and Spinoza 
Institute, Utrecht University,
   Utrecht, The Netherlands, \email{bdewit@phys.uu.nl}}
\author{J\"urg K\"appeli\\  Institute for Theoretical Physics and 
Spinoza Institute, Utrecht University,
   Utrecht, The Netherlands, \email{kaeppeli@phys.uu.nl}}
\author{Thomas Mohaupt\\Physics Department, Stanford
University, Stanford, CA 94305-4060, USA, \email{ mohaupt@itp.stanford.edu}}

\preprint{
SPIN-00/21\\[-2.5mm]
ITP-UU-00/24\\[-2.5mm]
SU-ITP 00/19\\[-2.5mm]
{\tt hep-th/0009234}}  

\abstract{ We analyze a broad class of stationary solutions with
residual $N=1$ supersymmetry of four-dimensional $N=2$ supergravity
theories with terms quadratic in the Weyl tensor. These terms are
encoded in a holomorphic function, which determines the  most relevant
part of the action and which plays a central 
role in our analysis. The solutions include extremal black
holes and rotating field configurations, and may have multiple
centers. We prove that they are expressed in terms of harmonic functions  
associated with the electric and magnetic charges carried by the
solutions by a proper generalization of the so-called stabilization 
equations. Electric/magnetic duality is manifest throughout the
analysis.
\\ 
We also prove 
that spacetimes with unbroken supersymmetry are fully determined
by electric and magnetic charges. This result  
establishes the so-called fixed-point behavior according to which the
moduli fields must flow towards certain prescribed values on a fully
supersymmetric horizon, but now in a more general context with
higher-order curvature interactions. We briefly comment on the implications 
of our results for the metric on the moduli
space of extremal black hole solutions. }

\begin{document}
\section{Introduction}\label{sec:intro}
\setcounter{equation}{0} 

In this paper we determine a broad class of stationary solutions 
of four-dimensional $N=2$ supergravity theories with
$R^2$-interactions. The supergravity theories that we consider are
based on vector multiplets and hypermultiplets coupled to the
supergravity fields and 
contain the standard Einstein-Hilbert action as well as terms
quadratic in the Weyl tensor. The most relevant part of the interactions
is encoded in a holomorphic function, which plays a central role in
our analysis. The solutions that we consider are BPS solutions,
because they possess a residual $N=1$ supersymmetry. Some of them describe
extremal black holes that carry electric and/or magnetic charges   
or superpositions thereof. We also describe rotating  
solutions with one or several centers. The extremal black holes are
solitonic interpolations between two fully supersymmetric 
groundstates. Without $R^2$-interactions these are flat 
Minkowski spacetime at spatial infinity and a Bertotti-Robinson
geometry at the horizon. In that case, the moduli fields, which can
take arbitrary values at infinity, must  
flow to specific values at the horizon which are determined in terms of
the charges. This so-called fixed-point behavior explains why the
black hole entropy depends only on the charges and not on the
asymptotic values of the moduli. This is in contradistinction with the
black hole mass which does depend on the values of the fields at
spatial infinity. Owing to this fixed-point behavior the resulting
expressions for the entropy, based on the effective low-energy action, can be
compared successfully with microstate counting results from string and
brane theory, which also depend exclusively on the charges.

Solutions based on supergravity actions without $R^2$-terms were analyzed some
time ago \cite{FKS}-\cite{Moore}. Important features of solutions with
$R^2$-interactions were presented more recently in a number of papers
\cite{BCDWLMS}-\cite{CDM2} and reviewed in \cite{mohaupt}.  In \cite{CDM} we
showed that corrections to the black hole entropy associated with $R^2$-terms
are in agreement with certain subleading corrections to the entropy (in the
limit of large charges) that follow from the counting of microstates
\cite{MSWV}. The main ingredients of the derivation in \cite{CDM} are the
behavior of the solution at the horizon and the use of a definition of the
black hole entropy that is appropriate when $R^2$-interactions are present. In
order to ensure the validity of the first law of black hole mechanics, we used
the definition provided by the Noether method of \cite{Wald}. This definition
leads to an entropy that deviates from the Bekenstein-Hawking area law.

The purpose of this work is to present the complete proof underlying the
results of \cite{CDM} and to further extend our study of solutions in the
presence of $R^2$-interactions. In particular we consider the full
interpolating extremal black hole solution, multi-centered solutions, as well
as general stationary solutions. All solutions known so far (e.g.
\cite{FKS}-\cite{Sabra}, \cite{Denef}) are contained as special cases. We begin our analysis
by determining all spacetimes with $N=2$ supersymmetry. In doing so we
systematize and complete the analysis presented in \cite{CDM}.  We prove that,
in spite of the presence of $R^2$-terms, there is still a unique spacetime,
which is of the Bertotti-Robinson type, whose radius as well as the values of
the various moduli fields are determined by the electric and magnetic charges
carried by the solution.  Flat Minkowski spacetime can be viewed as a special
case of such a solution, but here the moduli are constant and arbitrary and
there are no electric and magnetic fields. Our analysis thus shows that the
enhancement of supersymmetry at the horizon forces the moduli fields to take
prescribed values. Consequently the uniqueness of the horizon geometry implies
the existence of a fixed-point behavior even in the presence of
$R^2$-interactions. Note that the fixed-point behavior is usually derived by
invoking flow arguments based on the interpolating solutions (see, e.g.
\cite{FKS,Moore}), but these arguments are much more difficult to derive in
the presence of $R^2$-interactions.

Subsequently we turn to the analysis of spacetimes with residual
$N=1$ supersymmetry. 
A general analysis of the conditions for $N=1$ supersymmetry
turns out to be extremely cumbersome.  We therefore restrict ourselves to 
a well-defined class of embeddings of residual supersymmetry and
derive the corresponding restrictions on the bosonic background
configurations. Our analysis is set up in such a way that
the presence of the $R^2$-interactions hardly poses
complications. This is so 
because the $R^2$-terms are incorporated into the Lagrangian by
allowing the holomorphic function to depend on an
extra holomorphic parameter. Furthermore, by stressing the underlying
electric/magnetic duality of the field equations throughout the
calculations, the dependence on the $R^2$-interactions remains almost
entirely implicit and does not require much extra attention. 

Using the restrictions posed by residual supersymmetry and assuming
stationary field configurations we analyze the solutions. We prove
that they are expressed in terms of harmonic functions 
associated with the electric and magnetic charges carried by the
solutions, while the spatial dependence of the moduli is governed by 
so-called ``generalized stabilization equations''. The latter were first
conjectured in \cite{Sabra} and in \cite{BCDWLMS} for the case without and
with $R^2$-interactions, respectively.
The resulting stationary solutions include the case of multi-centered
solutions of extremal black holes.

Our analysis of the restrictions imposed by $N=2$ and $N=1$ supersymmetry
on the solutions is based on the existence of a full off-shell
(superconformal) multiplet calculus for $N=2$ supergravity theories 
\cite{DWVHVPL}. It turns out that the hypermultiplets play only a
rather passive role. It proves advantageous to perform
most of the analysis before writing the theory in its Poincar\'e form
(by imposing gauge conditions or reformulating it in terms of fields that
are invariant under the action of those superconformal symmetries that
are absent in Poincar\'e supergravity). As a consequence we fix the
stationary spacetime line element only at a relatively late stage of
the analysis. 
An unusual complication is that, in order
to determine the restrictions imposed by full or residual
supersymmetry, 
it is not sufficient to consider the supersymmetry variation of the 
fermions only. One also needs to impose the vanishing of the
supersymmetry  variation of derivatives
of the fermion fields.  We present an argument that shows which of these
variations are needed.

The paper is organized as follows.  In section \ref{sec:susy1} we
review the relevant features of the superconformal multiplet calculus
which we use to construct $N=2$ theories with $R^2$-interactions.  
We also briefly discuss electric/magnetic duality transformations in
the presence of  $R^2$-interactions.  In section \ref{sec:susy2} we
describe some of the technology needed for performing the analysis. In
particular, we construct various compensating fields for 
$S$-supersymmetry (which are inert with respect to
electric/magnetic duality) and we discuss a number of additional
transformation laws. In section \ref{sec:N=2} we perform a 
detailed analysis of the restrictions imposed by $N=2$ supersymmetry and 
make contact with previous results 
\cite{CDM}. Section \ref{sec:N=1} is devoted to the analysis of the 
restrictions imposed by $N=1$ supersymmetry for a particular class
of embeddings of residual supersymmetry. We derive the 
``generalized stabilization equations'' that determine the spatial
dependence of the moduli fields and prove that the solutions are
encoded in harmonic functions that are associated with the electric
and magnetic fields. Section \ref{sec:conclusions} contains our
conclusions as well as an outlook.   
Appendices \ref{a} and \ref{b} explain some of 
the definitions and conventions used throughout this paper.

\section{Action and supersymmetry transformation rules}\label{sec:susy1}
\setcounter{equation}{0}

The $N=2$ supergravity theories that we consider are based on abelian
vector multiplets and hypermultiplets coupled to the supergravity
fields. The action contains the standard Einstein-Hilbert action as
well as terms quadratic in the Riemann tensor. To describe such  
theories in a transparent way we make use of the superconformal
multiplet calculus 
\cite{DWVHVPL}, which incorporates the gauge symmetries of the
$N=2$ superconformal algebra. The corresponding high  
degree of symmetry allows for the use of relatively small    
field representations. One is the Weyl multiplet, whose fields 
comprise the gauge fields corresponding to the superconformal 
symmetries and a few auxiliary fields. The other ones are abelian
vector multiplets and hypermultiplets, as well as a general chiral
supermultiplet, which will be treated as independent in initial stages
of the analysis but at the end will be expressed in terms of the
fields of the Weyl multiplet.  
Some of the additional (matter) multiplets will provide
compensating fields which are necessary in order that the action
becomes gauge equivalent to a Poincar\'e supergravity theory. These
compensating fields bridge the deficit in degrees of freedom
between the Weyl multiplet and the Poincar\'e supergravity multiplet.  
For instance, the graviphoton, represented by
an abelian vector field in the Poincar\'e supergravity multiplet,
is provided by an $N=2$ superconformal vector multiplet. 

As we will demonstrate, it is possible to 
analyze the conditions for residual $N=1$ or full $N=2$ supersymmetry
directly in this superconformal setting, postponing a transition to
Poincar\'e supergravity till the end. This implies in particular that 
our intermediate results are subject to local 
scale transformations. Only towards the end we will convert to 
expressions that are scale invariant. 

The superconformal algebra contains general-coordinate, local Lorentz,
dilatation, special conformal, chiral SU(2) and U(1), supersymmetry 
($Q$) and special supersymmetry ($S$) transformations. The gauge 
fields associated with general-coordinate transformations ($e_\m^a$),
dilatations ($b_\m$), chiral symmetry (${\cal V}^{\;i}_{\m\, j}, A_\m$) 
and $Q$-supersymmetry ($\psi_\m^i$) are realized by independent
fields. The  remaining gauge fields of Lorentz ($\o^{ab}_\m$), special
conformal ($f^a_\m$) and $S$-supersymmetry transformations
($\phi_\m^i$) are dependent fields.  They are composite objects, which 
depend in a complicated way on the independent fields \cite{DWVHVPL}. 
The corresponding curvatures and covariant fields are contained
in a tensor chiral multiplet, which comprises $24+24$ off-shell 
degrees of freedom; 
in addition to the independent superconformal gauge fields it 
contains three auxiliary fields: a Majorana spinor doublet  
$\chi^i$, a scalar $D$ and a selfdual Lorentz tensor $T_{abij}$ (where $i,
j,\ldots$ are chiral SU(2) spinor indices)\footnote{%
  By an abuse of terminology, 
  $T_{abij}$ is often called the graviphoton field strength.
  It is antisymmetric in  both Lorentz indices $a,b$ and 
  chiral SU(2) indices $i,j$. Its complex conjugate is the anti-selfdual 
  field $T^{ij}_{ab}$. 
  Our conventions are such that SU(2) indices are 
raised and lowered by complex conjugation. The SU(2) gauge field 
${\cal V}_\m^{\;i}{}_j$ is antihermitian and traceless, i.e., 
${\cal V}_\m^{\;i}{}_j+{\cal V}_{\m j}{}^{i}= {\cal 
V}_\m^{\;i}{}_i=0$. }. %
We summarize the transformation rules for some of
the independent fields of the Weyl multiplet
under $Q$- and $S$-super\-symmetry and under 
special conformal transformations, with parameters $\e^i$, $\eta^i$ and
$\L^a_{\rm K}$, respectively,
\bea \d e_\mu{}^a &=&
     \bar{\e}^i\g^a\psi_{\mu i}+{\rm h.c.}\,, \nonumber\\
      \d\psi_\mu^i &=& 2{\cal D}_\mu\e^i
      -\ft18  T^{ab\,ij}\g_{ab}\,\g_\mu\e_j
      -\g_\mu\eta^i \,,\nonumber\\
      \d b_\mu &=&
      \ft12\bar{\e}^i\phi_{\mu i}
      -\ft34\bar{\e}^i\g_\mu\chi_i
      -\ft12\bar{\eta}^i\psi_{\mu i}+{\rm h.c.} +\L_{\rm K}^a\,e_{\m a}  
      \,,\nonumber\\   
      \d A_\mu &=& \ft{1}{2}i \bar{\e}^i\phi_{\mu i}
      +\ft{3}{4}i\bar{\e}^i\g_\mu\chi_i
      +\ft{1}{2}i \bar{\eta}^i\psi_{\mu i}+{\rm h.c.}\,, \nonumber\\
      \d T_{ab}^{ ij} &=&
      8 \bar{\e}^{[ i} {R}(Q)_{ab}^{j]}\,, \nonumber\\
      \d\chi^i &=& -\ft1{12}\g_{ab} D\!\llap/\, T^{ab\,ij}\e_j
      +\ft{1}{6} {R}({\cal V})_{ab}{}^{\!i}_{\;j}\,\g^{ab}\e^j
      -\ft{1}{3}i {R}({A})_{ab} \g^{ab} \e^i \nonumber\\
      && +D\,\e^i
      +\ft1{12} T^{ij}_{ab}\g^{ab}\eta_j \,,
 \label{transfo4}
\eea
where ${\cal D}_\mu$ are derivatives covariant with respect to
Lorentz, dilatational, U(1) and SU(2) transformations,
and $D_\mu$ are derivatives covariant with respect to all
superconformal transformations. Throughout this paper we suppress
terms of higher order in the fermions, as we will be dealing with
a bosonic background. The quantities ${R}(Q)^i_{\m\n}$,  
${R}(A)_{\m\n}$ and ${R}({\cal V})_{\m\n}{}^{\!i}_{\;j}$ 
are supercovariant curvatures related to $Q$-supersymmetry, U(1) and
SU(2) transformations. Their precise definitions are given in
appendix \ref{b}. We will make explicit use of the transformation of
the $S$-supersymmetry gauge field, 
\beq
\d\phi^i_\m = - 2\,f_\m^a\g_a\e^i + \ft14 R({\cal V})_{ab}{}^{\,
i}{}_{\!j} 
\g^{ab}\g_\m \e^j + \ft12i  R(A)_{ab}\g^{ab}\g_\m \e^i-\ft18 
D\!\llap/\, T^{ab\,ij} \g_{ab} \g_\m \e_j +  2{\cal D}_\m\eta^i\,.
\label{phi-transfo}
\eeq
Here $f_\m^{\,a}$ is the gauge field of the conformal boosts, 
defined by (up to fermionic terms)
\beq
f_\m^{\,a}= \ft12 R(\omega,e)_{\m}{}^a - \ft14 (D+ \ft13 R(\omega,e)) \,
e_\m^{\,a}  - \ft12 i 
\tilde R_{\m}{}^a(A) + \ft1{16} T_{\m b}^{ij}\, T^{ab}_{ij} \,,
\label{f-def}
\eeq
with $R(\omega,e)_{\m}{}^{a}= R(\omega)_{\m\n}^{ab} \,e_b^\n$ the
(nonsymmetric)  
Ricci tensor and $R(\omega,e)$ the Ricci scalar. Here
$R(\omega)_{\m\n}^{ab}$ is the curvature associated with the spin
connection field $\omega_\m^{ab}$. The spin connection is not the usual
torsion-free connection, but contains the dilatational gauge field
$b_\m$. Because of that the curvature satisfies the Bianchi identity 
\beq
R(\omega)_{[\m\n}^{ab} \,e_{\rho]\,b} = 2 \pa_{[\m}b_{\n}
\,e_{\rho]}^{\,a} \,.  
\ee
This leads to the modified pair-exchange property 
\beq
R(\omega)_{ab}{}^{\!cd} -R(\omega)^{cd}{}_{\!ab} = 2 \Big
( \d^{[c}_{[a} R(\omega,e)_{b]}{}^{d]} -   
\d^{[c}_{[a} R(\omega,e)^{d]}{}_{b]}\Big)  \,.
\eeq

It is important to mention that the covariant quantities of the Weyl
multiplet constitute a reduced chiral tensor multiplet, denoted by
$W^{abij}$, whose lowest-$\theta$ component is the tensor
$T^{abij}$. In this multiplet the superconformal gauge fields appear 
only through covariant derivatives and curvature tensors. 
{}From this multiplet one may form a scalar (unreduced) chiral
multiplet $W^2 = [W^{abij}\,\varepsilon_{ij}]^2$, which has Weyl and
chiral weights $w=2$ and $c=-2$, respectively \cite{BDRDW}.  

In the following, we will allow for the presence of an arbitrary
chiral background superfield \cite{deWit}, whose component fields will
be indicated with a caret.  Eventually this multiplet will be
identified with $W^2$ in order to generate the $R^2$-terms in the
action, but much of our analysis will not depend on this
identification.  We denote its bosonic component fields by ${\hat A}$,
${\hat B}_{ij}$, ${\hat F}_{ab}^-$ and by ${\hat C}$.  Here ${\hat A}$
and ${\hat C}$ denote complex scalar fields, appearing at the
$\theta^0$- and $\theta^4$-level of the chiral background superfield,
respectively, while the symmetric complex SU(2) tensor ${\hat B}_{ij}$
and the anti-selfdual Lorentz tensor ${\hat F}_{ab}^-$ reside at the
$\theta^2$-level.  The fermion fields at level $\theta$ and $\theta^3$
are denoted by $\hat\Psi_i$ and $\hat\Lambda_i$. Under $Q$- and
$S$-supersymmetry $\hat A$ and $\Psi_i$ transform as \beqa \d {\hat A}
&=& \bar{\e}^i {\hat \Psi}_i \,,\nonumber\\ \d {\hat \Psi}_i &=& 2
D\!\llap/\, {\hat A} \e_i +\ft12 \ve_{ij} {\hat F}_{ab} \g^{ab} \e^j
+{\hat B}_{ij} \e^j +2 w {\hat A} \eta_i\,, \eeqa where $w$ denotes
the Weyl weight of the background superfield. 
Identifying the scalar chiral multiplet with $W^2$ implies the
following relations, which we will need later on, 
\bea
\hat A   &=& (\varepsilon_{ij}\,T^{ij}_{ab})^2\,,\nonumber \\
\hat \Psi_i &=& 16\, \varepsilon_{ij}R(Q)^j_{ab} \,T^{klab} \,
\varepsilon_{kl} \,,\nonumber\\  
\hat B_{ij}  &=& -16 \,\varepsilon_{k(i}R({\cal V})^k{}_{j)ab} \,
T^{lmab}\,\varepsilon_{lm} -64 \,\varepsilon_{ik}\varepsilon_{jl} 
\bar R(Q)^k_{ab} R(Q)^{l\,ab}   \,,\nonumber\\
\hat F^{-ab}  &=& -16 \,{\cal R}(M)_{cd}{}^{\!ab} \,
T^{klcd}\,\varepsilon_{kl}  -16 \,\varepsilon_{ij}\, \bar R(Q)^i_{cd} 
\gamma^{ab} R(Q)^{j\,cd}  \,,\nonumber\\
\hat \Lambda_i &=&32\, \varepsilon_{ij} \,\g^{ab} R(Q)_{cd}^j\, 
{\cal R}(M)^{cd}{}_{\!ab} 
+16\,({\cal R}(S)_{ab\,i} +3 \g_{[a} D_{b]}  \chi_i) \, 
T^{klab}\, \varepsilon_{kl} \nonumber\\
&& -64\, R({\cal V})_{ab}{}^{\!k}{}_i \,\varepsilon_{kl}\,R(Q)^{l\,ab}
\,,\nonumber\\ 
\hat C &=&  64\, {\cal R}(M)^{-cd}{}_{\!ab}\, {\cal 
R}(M)^-_{cd}{}^{\!ab}  + 32\, R({\cal V})^{-ab\,k}{}_l^{~} \, 
R({\cal V})^-_{ab}{}^{\!l}{}_k^{~} \nonumber \\
&& - 32\, T^{ab\,ij} \, D_a \,D^cT_{cb\,ij} +128 \, \bar {\cal 
R}(S)^{ab}_i \,R(Q)_{ab}^i  +384 \,\bar R(Q)^{ab\,i} 
\gamma_aD_b\chi_i   \,.   
\label{background-def}
\eea
We refer to appendix \ref{b} for the definitions of the various 
curvature tensors. 

Let us briefly introduce the hypermultiplets, which play a rather
passive role in what follows but are needed to provide one of the
compensating supermultiplets. Here we follow the presentation of
\cite{DWKV}, based on sections $A_i{}^\a(\phi)$ which depend on scalar
fields $\phi^A$, defined in the context of a so-called special
hyper-K\"ahler space, endowed with a metric $g_{AB}$, and a
tangent-space  connection
$\Gamma_A{}^{\!\a}{}_\b$ as well as two covariantly
constant tensors, $\bar\Omega_{\a\b}$ and $G_{\a\bar\b}$, which are
skew-symmetric pseudo-real and hermitian, respectively. The positive
and negative chirality fermions are denoted by $\zeta^{\bar\a}$ and 
$\zeta^\a$ and are related by complex conjugation. The indices $\a$ and
$\bar \a$ run over $2r$ values while 
the number of scalar fields labeled by indices $A$ is equal to
$4r$. Hence the special hyper-K\"ahler space has dimension $4r$, while
the number of physical hypermultiplets will be given by $r-1$. For
what follows, it suffices to consider the variations of the 
fermion fields $\zeta^\a$ under $Q$- and $S$-supersymmetry
transformations,   
\beq
\d \zeta^\a = D\!\llap/ \,A_i{}^\a \e^i- 
\d \phi^B \G_B{}^{\!\a}{}_{\!\b}\, \zeta^\b 
 + A_i{}^\a\,\eta^i\,.
\label{hypertransf} 
\eeq
Here we have assumed that the hypermultiplets are neutral with
respect to the gauge symmetries of the vector multiplets (to be
introduced below), so there is no minimal interaction with the vector
multiplet fields. The bosonic part of $D_\m A_i{}^\a(\phi)$
will be given shortly. 

Finally we turn to the abelian vector multiplets, labelled by an 
index $I= 0,1,\ldots,n$. For each value of the index $I$,
there are $8+8$ off-shell degrees of freedom, residing in a 
complex scalar $X^{I}$, a doublet of chiral fermions 
$\Omega_i^{\,I}$, a vector gauge field $W_\mu^{\,I}$,
and a real SU(2) triplet of auxiliary scalars $Y_{ij}^{\,I}$. 
Under $Q$- and $S$-supersymmetry the fields $X^I$ and $\Omega_i^{\,I}$
 transform as follows,
\bea 
\d X^{I} &=& \bar{\e}^i\Omega_i^{\,I} \,,\nonumber\\
     \d\Omega_i^{\,I} &=& 2 D\!\llap/\, X^{I}\e_i
     +\ft12 \ve_{ij} ( F^{-I}_{\m\n} -\ft14
   \ve_{kl}T^{kl}_{\mu\nu}\,\bar{X}^{I})   \g^{\m \n} \e^j
     +Y_{ij}^{\,I}\e^j
     +2X^{I}\eta_i\,,
\label{vrules}
\eea
where $F_{\mu\nu}^{\pm\,I}$ are the (anti-)selfdual parts of the vector
field strength, $F_{\mu\nu}^{+I}+ F_{\mu\nu}^{-I}=
2\pa_{[\mu}W_{\nu]}^I$. 

The covariant quantities of the vector multiplet constitute a 
reduced chiral multiplet whose lowest component is the complex scalar
$X^I$, which has Weyl and chiral weights $w=1$ and $c=-1$,
respectively. A general 
(scalar) chiral multiplet comprises $16+16$ off-shell degrees of freedom
and carries arbitrary Weyl and chiral weights. 
The supersymmetric action is now constructed from a chiral superspace
integral of a holomorphic function of these reduced chiral
multiplets. However, in order to preserve the superconformal
symmetries this function must be homogeneous of second
degree. This implies that its weights are $w=2$ and $c=-2$. An
important observation is that this function can depend on any other
chiral field, as long as its scale and chiral weights are properly
accounted for. In particular, this means that we can base ourselves on
a homogeneous function $F(X,\hat A)$ which is of degree two, that
depends on the complex fields $X^I$ and on the scalar of the
background chiral multiplet, $\hat A$. Therefore this function
satisfies the relation,
\beq 
X^I F_I + w \hat A \,F_A = 2 F\,.
\eeq
Here $F_I$ and $F_A$ denote the derivatives of $F(X,\hat A)$ with
respect to $X^I$ and $\hat A$, respectively, and $w$
denotes the weight of the background field.

In the absence of a background it is known that there are
representations of the theory for which no function $F(X)$ exists,
although after a suitable electric/magnetic duality transformation 
it can be rewritten in a form that exhibits the function $F(X)$. In
the presence of a background, this feature does not seem to play a
direct role, so we will simply 
assume the existence of $F(X,\hat A)$. For 
some of the notations and background material, see \cite{deWit} and the 
third reference of \cite{DWVHVPL}, where a general chiral 
multiplet in supergravity is discussed.

The bosonic terms of the action are encoded in the function
$F(X,\hat A)$, in the hypermultiplet sections $A_i{}^\a(\phi)$ and in the
target space connections. They read as follows,
\bea
\label{efflag}
8\pi\,e^{-1}\, {\cal L} &=&  
 i {\cal D}^{\mu} F_I \, {\cal D}_{\mu} \bar X^I   - i F_I\,\bar X^I 
 (\ft16  R - D) 
-\ft18i  F_{IJ}\, Y^I_{ij} Y^{Jij} - \ft14 i \hat 
B_{ij}\,F_{{ A}I}  Y^{Iij}   \nonumber\\
&&+\ft14 i F_{IJ} (F^{-I}_{ab} -\ft 14 \bar X^I 
T_{ab}^{ij}\varepsilon_{ij})(F^{-Jab} -\ft14 \bar X^J 
T^{ijab}\varepsilon_{ij})  \nonumber\\
&&-\ft18 i F_I(F^{+I}_{ab} -\ft14  X^I 
T_{abij}\varepsilon^{ij}) T^{ab}_{ij}\varepsilon^{ij}  
+\ft12 i \hat F^{-ab}\, F_{{ A}I} (F^{-I}_{ab} - \ft14  \bar X^I 
T_{ab}^{ij}\varepsilon_{ij})   \nonumber \\
&&+\ft12 i F_{A}
\hat C -\ft18 i F_{ A A}(\varepsilon^{ik}
\varepsilon^{jl}  \hat B_{ij} 
\hat B_{kl} -2 \hat F^-_{ab}\hat F^{-ab}) 
-\ft1{32} iF (T_{abij}\varepsilon^{ij})^2 + {\rm h.c.}\nonumber \\ 
&& - \ft12 \varepsilon^{ij}\, \bar \Omega_{\a\b} \,
{\cal D}_\m A_i{}^\a \,{\cal D}^\m  A_j{}^\b +\chi (\ft16 R+  \ft12 D) \;,
\eea
where the hyper-K\"ahler potential  $\chi$ and the covariant
derivative ${\cal D}_\m A_i{}^\a(\phi)$ are defined by
\bea
\varepsilon_{ij}\,\chi &=& \bar\Omega_{\a\b} A_i{}^\a A_j{}^\b
\,,\nonumber\\  
{\cal D}_\m A_i{}^\a &=& \partial_\m A_i{}^\a -
b_\m A_i{}^\a +\ft12V_{\m i}{}^jA_j{}^\a  +  
\pa_\m\phi^A\,{\G_A}^{\!\a}{}_{\!\b} \, A_i{}^\b\,.
\label{hyperKpotential}
\eea

Even in the presence of the chiral background the Lagrangian has the
form of a generalized Maxwell Lagrangian with terms that are at most
quadratic in the field strengths. This feature will change once we
start eliminating auxiliary fields.\footnote{
  Because the chiral background field given in \eqn{background-def} 
  involves terms of higher order in derivatives, the Lagrangian  will
  contain higher-derivative interactions. The most conspicuous
  ones are the interactions quadratic in the Riemann curvature. Such
  Lagrangians generically describe negative-metric states. However,
  they should not be regarded as elementary Lagrangians, but rather as
  effective Lagrangians. This implies that auxiliary fields that appear
  with derivatives, should still be eliminated. This leads to 
  an infinite series of terms that corresponds to an expansion in
  terms of momenta divided by the Planck mass. 
}  
Hence it is advisable to first
solve the Maxwell equations, before eliminating the auxiliary
fields. One distinguishes the Bianchi equations, which  are
expressed directly in terms of the field strengths $F_{\m\n}^{\pm\,I}$,
and the equations for the electric and magnetic `displacement' fields
$G^\pm_{\m\n I}$, which are
proportional to the variation of the action with respect to the
$F_{\m\n}^{\pm\,I}$. With suitable proportionality factors, these
tensors read (we suppress fermion contributions), 
\beqa
G^+_{\mu\nu I}= \bar F _{IJ}F^{+J}_{\mu\nu} + {\cal O}_{\mu\nu 
I}^+\,, \qquad G^-_{\mu\nu I}= F_{IJ}F^{-J}_{\mu\nu} + 
{\cal O}_{\mu\nu I}^- \,, \label{defG}
\eeqa
where 
\bea
{\cal O}_{\mu\nu I}^+   &=&\ft14 (F_I-\bar F_{IJ}X^J )\,T_{\mu\nu 
ij}\varepsilon^{ij} +\hat F^+_{\mu\nu} \,\bar  
F_{I{A}} \,,\nonumber \\ 
{\cal O}_{\mu\nu I}^-   &=& \ft14 (\bar F_I- F_{IJ}\bar X^J
)\,T_{\mu\nu}^{ij}\, 
\varepsilon_{ij} +\hat F^-_{\mu\nu} \, F_{I{A}} \,. \label{NO} 
\eea
In terms of these tensors the Maxwell equations in the absence of
charges read (in the presence of the background), ${\cal D}^a ( F^- -
F^+)_{ab}^I = 0$, and  ${\cal D}^a (G^- - G^+)_{ab\,I} =0$. Eventually
we will solve 
these equations for a given configuration of electric and magnetic
charges in a stationary geometry. These charges will be denoted by
$(p^I,q_J)$ and are normalized such that for a stationary multi-centered
solution with charges at centers $\vec{x}_A$, Maxwell's equations read
\begin{equation}
  \label{eq:normcharges}
    \partial_\mu \pmatrix{ \sqrt{g} (F^{-} - F^{+})^{I\,\mu
  t}\cr\noalign{\vskip2mm}  \sqrt{g} ( G^{-} - G^{+})^{\mu t}_I}  = 4 i \pi \sum_A
   \delta(\vec{x} - \vec{x}_A)\pmatrix{p^I_A\cr\noalign{\vskip2mm} q_{AI}} \,.
\end{equation}
Observe that $ \sqrt{g}\, (F^{-} - F^{+})^{I\,\mu
  \nu}$ and  $ \sqrt{g}\, ( G^{-} - G^{+})^{\mu \nu}_I $ are Weyl invariant
  quantities. 

The field equations of the vector multiplets are subject to 
equivalence transformations corresponding to electric/magnetic 
duality, which do not affect the fields of the Weyl multiplet
and of the chiral background. As is well-known, the following two complex
$(2n+2)$-component vectors transform linearly
under the SP$(2n+2;{\bf R})$  duality group,
\beqa
\pmatrix{X^I\cr\noalign{\vskip1mm} F_I(X,{\hat A})} \quad 
{\mbox{and}} \quad
\pmatrix{F_{ab}^{\pm\,I}\cr\noalign{\vskip1mm} G^\pm_{ab\,I}} \;,
\label{seca} 
\eeqa 
but more such vectors can be constructed.  The first vector has weights
$w=1$ and $c=-1$, whereas the second one has zero Weyl and chiral weights.
From \eqn{eq:normcharges} and \eqn{seca} it follows that also the charges
$(p^I,q_J)$ comprise a symplectic vector. In the presence of these
charges the symplectic transformations are restricted to an integer-valued
subgroup that keeps the lattice of electric/magnetic charges invariant.

The electric/magnetic duality transformations cannot be performed at the
level of the action, but only at the level of the equations of
motion. After applying the transformations one can find the
corresponding action. This is then characterized by a relation between
two different functions $F(X,\hat A)$. While the
background field $\hat A$ is inert under the dualities, it 
nevertheless enters in the explicit form of the transformations. For a
discussion of this phenomenon and its consequences, see \cite{deWit}. 

The various transformation rules only take a symplectically invariant
form when one solves the field equations for the auxiliary fields
$Y^I_{ij}$ \cite{deWit}, 
\beq 
Y_{ij}^I= i N^{IJ} \Big(F_{JA}\, \hat B_{ij} - \bar F_{JA}\, 
\varepsilon_{ik}\varepsilon_{jl}\, \hat B^{kl} \Big)\,.
\eeq
With this result we can cast $\d\Omega_i^I$ and $\d\hat\Psi_i$ in a
symplectically covariant form (we suppress fermionic bilinears), 
\bea
&&\pmatrix{\d\O^I_i\cr \noalign{\vskip 1mm} \d(F_{IJ} \O^J_i +
F_{IA}\hat\Psi_i)\cr} 
= 2 D\!\llap/\, \pmatrix{X^{I}\cr \noalign{\vskip 1mm} F_I\cr} \e_i
     +\ft12 \ve_{ij} \g^{ab} \e^j\, \left[ \pmatrix{F_{ab}^{-I}\cr
\noalign{\vskip 1mm}  
G_{abI}^-\cr}  - \ft14 \varepsilon_{kl} T^{kl}_{ab} \,\pmatrix{\bar
X^I\cr \noalign{\vskip 1mm} \bar F_I\cr} \right]  \nonumber \\[3mm]
&& + i \hat B_{ij}\,\e^j \pmatrix{N^{IJ} F_{JA}\cr \noalign{\vskip
1mm}  \bar F_{IJ} N^{JK} F_{KA}
\cr} - i \varepsilon_{ik} \varepsilon_{jl}\hat B^{kl} \,\e^j \, 
\pmatrix{N^{IJ}
\bar F_{JA} \cr \noalign{\vskip 1mm} F_{IJ} N^{JK} \bar F_{KA} \cr} +
2\eta_i\, \pmatrix{ X^{I}\cr \noalign{\vskip 1mm} F_I\cr} \,.
\label{symplvectortransform}
\eea
In the above formulae, $N^{IJ}$ is the inverse of 
\beq
N_{IJ} = -i F_{IJ} + i \bar F_{IJ} \,.
\eeq
\section{More supersymmetry variations}\label{sec:susy2}
\setcounter{equation}{0}

In the superconformal tensor calculus two of the matter
supermultiplets are required in order to provide the compensating
degrees of freedom that are essential for making the system equivalent
to a Poincar\'e supergravity theory. One of these multiplets is always
a vector multiplet and for the second one we choose a
hypermultiplet. This implies that the number of physical vector
multiplets is equal to $n$ and the number of physical hypermultiplets
is equal to $r-1$. 

In this section we will evaluate the supersymmetry variations of a
number of spinors that are needed in the analysis in subsequent
sections. The results of this section follow from those
given in the previous one. Some of the spinors can act as suitable
compensating fields with regard to $S$-supersymmetry. We also evaluate 
the supersymmetry 
variations of the supercovariant derivative of the spinors belonging
to one of the matter multiplets as well as the variation of the
supersymmetry field strength $R(Q)^i_{ab}$. This analysis naturally
leads us to the definition of a number of bosonic quantities that play
a central role in what follows.

The first spinor we consider is expressed in terms of hypermultiplet
fermions and reads 
\beq
\zeta_i^{\scriptscriptstyle\rm H} \equiv \chi^{-1}
\bar\Omega_{\a\b} A_i{}^\a \,\zeta^\b \,.
\eeq
Its supersymmetry variation reads
\bea
\d\zeta^{\scriptscriptstyle\rm H}_i = \chi^{-1} \bar\Omega_{\a\b}
A_i{}^\a  D\!\llap/\,A_j{}^\b \,\e^j + \varepsilon_{ij} \eta^j \,, 
\eea
where $\chi$ is the hyper-K\"ahler potential defined in
\eqn{hyperKpotential} and where terms proportional to
the fermion fields are suppressed. It is important to realize that
one has the decomposition \cite{DWKV}
\beq
\chi^{-1} \bar\Omega_{\a\b} A_i{}^\a D_\m A_j{}^\b  = \ft12
 k_\m \,\varepsilon_{ij} + k_{\m\,ij}\,,
\eeq    
where $k_\m$ is  real and given by
\beq
k_\m = \chi^{-1} (\pa_\m -2 \,b_\m)\chi\,,
\eeq
and $k_{\m\,ij}$ is symmetric in
$i,j$ and pseudoreal so that it transforms as a vector under
SU(2). Its explicit form is not important for us. Hence we
write 
\bea
\d\zeta^{\scriptscriptstyle\rm H}_i = \ft12
k\!\llap/\,\,\varepsilon_{ij}\, \e^j + k\!\llap/\,_{ij}\,\e^j  
 + \varepsilon_{ij}\, \eta^j \,. 
\eea

In the vector multiplet sector there are two spinors that can be
constructed which transform as scalars under electric/magnetic
duality. One, denoted by $\zeta_i^{\scriptscriptstyle V}$, transforms
inhomogeneously under $S$-supersymmetry.  It can be conveniently
introduced from the variation of the symplectically 
invariant expression  (with $w=2$ and $c=0$)
\beq
{\rm e}^{-{\cal K}} = i\Big[\bar X^I\, F_I(X,\hat A) - \bar F_I(\bar X,
\bar{\hat A}) \, X^I \Big] \,.
\label{kaehler}
\eeq
Here $\cal K$ resembles the
K\"ahler potential in  special geometry. Its supersymmetry variation
leads to the spinor 
\bea
\zeta_i^{\scriptscriptstyle \rm V}  \equiv - \Big(\Omega^I_i
\,{\pa\over\pa X^I} + \hat\Psi_i \, 
{\pa\over\pa \hat A}\Big) {\cal K} 
=-i \,{\rm e}^{\cal K}\Big[  (\bar F_I - \bar 
X^JF_{IJ}) \Omega_i^I  -  \bar X^I F_{I{ A}} \, 
\hat\Psi_i\Big]\,.
\eea
It is obvious that $\zeta_i^{\scriptscriptstyle \rm V}$ transforms as
a scalar under symplectic reparameterizations, because it follows from a
symplectic 
scalar. This can also been seen 
by noting that $\zeta_i^{\scriptscriptstyle \rm V}$ is generated by
the symplectic product $\bar F_I \,\d X^I - \bar X^I\,\d F_I$. This
leads us to introduce yet another spinor $\zeta_i^{\scriptscriptstyle
0}$ generated by $F_I \,\d X^I -  X^I\,\d F_I$,
\beq
\zeta_i^{\scriptscriptstyle 0}  \equiv  (F_I - X^J F_{IJ}) \Omega_i^I
-  X^I F_{I{A}} \,  \hat\Psi_i  \,.
\eeq
This spinor is 
invariant under $S$-supersymmetry and it vanishes in the absence of
the chiral scalar background field. However, it does not play a useful
role in what follows. 

Under $Q$- and $S$-supersymmetry $\zeta_i^{\scriptscriptstyle \rm V}$
transforms as  
\bea
\d\zeta_i^{\scriptscriptstyle \rm V}  &=&   {\rm e}^{\cal K}
{\cal D}\!\llap/\, {\rm e}^{-\cal K}\e_i  
 +2 i{\cal A}\!\llap/\,\, \e_i  - \ft12 i 
\ve_{ij} \, {\cal F}_{ab}^- \, \g^{ab} \e^j \nonumber \\
&& + {\rm e}^{\cal K} \,N^{IJ}\Big[(\bar F_I -  \bar F_{IK}\bar X^K )
F_{JA} \, \hat B_{ij}  - (\bar F_I -  F_{IK}\bar X^K ) \bar 
F_{JA} \,\varepsilon_{ik} 
\varepsilon_{jl} \hat B^{kl} \Big] \, \e^j 
 +2  \,\eta_i\,,
\label{susyzeta}
\eea
where we ignored higher-order fermionic terms.
The quantity ${\cal A}_\m$ resembles a covariantized (real) K\"ahler
connection and ${\cal F}_{ab}^-$ is an anti-selfdual tensor,
\beqa
{\cal A}_\m &=& \ft12 {\rm e}^{\cal K}
\Big( \bar X^J \stackrel{\leftrightarrow}{\cal D}_\m  F_{J} 
-\bar F_J \stackrel{\leftrightarrow}{\cal
D}_\m X^{J} \Big) \,,\nonumber \\
{\cal F}_{ab}^- &=& {\rm e}^{{\cal K}} \, 
 \left( \bar F_I \,F^{-I}_{ab} - 
\bar X^I \,G_{ab\,I}^- \right) \;\;.
\label{curlyf}
\eeqa
There is another symplectically invariant contraction of the field
strengths,
\beq
{\rm e}^{{\cal K}} \, 
 \left(  F_I \,F^{-I}_{ab} -  X^I \,G_{ab\,I}^- \right) + \ft14 i
\varepsilon_{ij}  
T_{ab}^{ij}  = {\rm e}^{{\cal K}} \, F_{IA} \Big[w\hat A(
F_{ab}^{-I} -\ft14 \bar X^I \,\varepsilon_{ij} T^{ij}_{ab} ) -
X^I\,\hat F_{ab} ^-\Big] \,,
\eeq
which appears in the variation of $\zeta_i^{\scriptscriptstyle 0}$. 

As it turns out we also need to consider the supersymmetry variations
of derivatives of the fermion fields. However, one can restrict
oneself to the variation of the supercovariant derivative of
a single fermion field, as we will discuss in the next section. For
this field we choose $\zeta^{\scriptscriptstyle\rm H}_i$, for which we
present the variation under $Q$- and $S$-supersymmetry, 
\bea
\d( D_\m \zeta^{\scriptscriptstyle\rm H}_i) &=& \ft12
{\cal D}_\m (\chi^{-1} {\cal D}_\n\chi)\,\varepsilon_{ij}\,\g^\n \e^j
+ {\cal D}_\m 
k_{\n ij}\,\g^\n \e^j    \nonumber \\
&&-  \ft 1{32} \chi^{-1/2} (\d_i^j {\cal D}_\n  - k_{\n
ik}\varepsilon^{kj})(\chi^{1/2} T^{lm}_{ab}  
\varepsilon_{lm} )\, \g^\n\g^{ab} \g_\m \e_j  \nonumber \\ 
&&+  \varepsilon_{ij}\, \Big[ f_\m{}^a\g_a \e^j 
-\ft18 R({\cal V})^j{}_{k ab}   
\g^{ab}\g_\m\e^k - \ft14 i R(A)_{ab} \g^{ab} \g_\m \e^j \Big]
\nonumber \\
&&+ (\ft14 \chi^{-1}{\cal D}\!\llap/\,\chi\,\varepsilon_{ij} + \ft12
k\!\llap/\,_{ij})\,\g_\m \eta^j  \,.  
\eea
 
Finally we present the variation of the curvature tensor
$R(Q)_{\m\n}^i$, defined by 
\bea 
{R}(Q)_{\mu\nu}^i =
     2{\cal D}_{[\mu}\psi_{\nu]}^i
     -\g_{[\mu}\phi_{\nu]}^i
     -\ft{1}{8} T^{ij}_{ab}\g^{ab} \g_{[\mu}\psi_{\nu]j}\,,
\eea
where $\phi_\m^i$ is the dependent gauge field associated with 
$S$-supersymmetry, defined in appendix \ref{b}. 
The variation of this tensor reads,
\bea 
\d R(Q)_{ab}^i &=& - \ft12 D\!\llap/\, T^{ij}_{ab} \,\e_j +
R({\cal V})_{ab}^-{}^i{}_j \, \e^j \nonumber \\
&& - \ft12 {\cal R}(M)_{ab}{}^{\!cd}\, \g_{cd} \e^i +\ft18  
T^{ij}_{cd}\,\g^{cd}  \g_{ab} \,\eta_j  \,,
\eea
where ${\cal R}(M)_{ab}{}^{\!cd}$ is defined in appendix \ref{b}. 

\section{Fully supersymmetric field configurations}\label{sec:N=2}
\setcounter{equation}{0}
{From} the supersymmetry variations presented in the previous two
sections one can determine the conditions on the bosonic 
fields imposed by the requirement of full $N=2$ supersymmetry. These
conditions follow from setting all $Q$-supersymmetry variations of the
fermionic quantities to 
zero. However, these variations are determined up to an $S$-supersymmetry
transformation. Thus one can either impose the vanishing of all
$Q$-variations up to a uniform $S$-supersymmetry transformation, or
one can restrict oneself to 
linear combinations that are invariant under $S$-supersymmetry and
require their $Q$-supersymmetry variations to vanish. Examples of such
$S$-invariant combinations are, for instance,
$\O^I_i-X^I\zeta_i^{\scriptscriptstyle \rm V}$ and
$\hat \Psi_i - w \hat A\, \zeta_i^{\scriptscriptstyle \rm V} $, while
the spinor $\zeta_i^{\scriptscriptstyle \rm 0}$ is $S$-invariant by
itself. In this section we will include an arbitrary number of
hypermultiplets. 

We start by considering the $S$-supersymmetric linear combination of 
$\zeta_i^{\scriptscriptstyle \rm V}$ and $\zeta_i^{\scriptscriptstyle
\rm H}$. Requiring its $Q$-supersymmetry variation to vanish for all 
supersymmetry parameters, we establish immediately that
\beq
{\cal F}^-_{ab}  = \hat B_{ij}  = k_{\m\,ij}={\cal A}_\m = 0\,,
\eeq
and
\beq
{\cal D}_\m \Big({\rm e}^{{\cal K}}  \chi\Big) = 0\,. 
\label{chiK}
\eeq
Comparing the supersymmetry variations of the vector multiplet
fermions to those of $\zeta_i^{\scriptscriptstyle \rm V}$ leads to
\bea
F_{ab}^{-I} &=&  \ft14 \varepsilon_{kl} T^{kl}_{ab} \,\bar X^I\,,
\nonumber \\
G_{abI}^- &=&   \ft14 \varepsilon_{kl} T^{kl}_{ab} \,\bar F_I\,,
\nonumber \\
{\cal D}_\m \Big({\rm e}^{{\cal K}/2} X^I\Big) &=&  {\cal D}_\m
\Big({\rm e}^{{\cal K}/2} F_I\Big)  = 0\,. 
\label{XK}
\eea
These equations themselves again imply that ${\cal F}^-_{ab}$
and ${\cal A}_\m$ vanish. Furthermore, by using the explicit
expression of the tensors $G^-_{abI}$, one finds that $\hat
F^-_{ab}=0$. The last two equations imply that we also have 
\beq
{\cal D}_\m \Big({\rm e}^{w {\cal K}/2} \hat A\Big) =0\,. 
\eeq
{From} the supersymmetry variations of the hypermultiplets we find a
similar result,
\bea
{\cal D}_\m \Big(\chi^{-1/2} A_i{}^\a\Big) = 0\,.
\label{AK}
\eea
Observe that all the above equations are $K$-invariant. 

Subsequently we compare the supersymmetry variations of the spinors
$\chi^i$ and $\zeta_i^{\scriptscriptstyle \rm V}$, which leads to the
relations,
\bea
D=R({\cal V})_{ab}{}^{\!i}{}_j = R(A)_{ab} = {\cal D}_a \Big({\rm
e}^{-{\cal K}/2} T^{abij} \Big) = 0\,.  
\eea
With these results it follows that the vector field strengths satisfy
the following equations,
\beq
{\cal D}^a F^{-I}_{ab} =  {\cal D}^a G^{-}_{abI} =  0\,,
\eeq
which imply (but are stronger than) the equations of motion and the
Bianchi identities for the vector fields.

A similar calculation for the curvature $R(Q)_{ab}^i$ yields
\bea
{\cal D}_c T_{ab}^{ij} &=& -\ft12 {\cal D}_d {\cal K} \,\Big( \d^d_c\,
T^{ij}_{ab} 
- 2\d^d_{[a} \,T^{ij}_{b]c} + 2 \eta_{c[a}\,T^{ij\,d}_{\,b]}\Big) \,,  
\nonumber \\    
{\cal R}(M)_{ab}{}^{\!cd} &=& 0\,.
\eea
The first equation is consistent with the result found
earlier. Because $D=0$, ${\cal R}(M)_{ab}{}^{\!cd}$ is just the
traceless part 
of the curvature tensor $R(\omega)_{ab}{}^{\!cd}$ associated with the
spin connection field $\omega_\m^{ab}$ (which  at this stage 
depends on the dilatational gauge field $b_\m$).
Upon suppressing $b_\m$, this tensor becomes equal to the Weyl
tensor. Hence the 
above condition will eventually lead to the conclusion that $N=2$
supersymmetric solutions require a conformally flat spacetime. We
stress again that all of the above conditions are $K$-invariant. 

Before continuing, let us make a few remarks. First of all, we note
that at this stage all equations are consistent with all the
superconformal symmetries; in particular, we have not yet fixed a
scale. All the above results are also manifestly consistent with
electric/magnetic duality. Secondly we found a number of conditions on
the chiral 
background field, namely $\hat B_{ij} = \hat F_{ab}^-=0$ and the
covariant constancy of $\exp(w{\cal K}/2) \hat A$. So far no conditions
have been derived for its highest-$\theta$ component $\hat C$, but by
considering the supersymmetry variation of the spinor $\hat \Lambda_i$
one can easily show that $\hat C=0$. It is illuminating
to verify whether these results hold for the chiral field starting
with $\hat A= [T^{abij}\varepsilon_{ij}]^2$. It turns out that they
are indeed satisfied on the basis of the above results, with the
exception of the $\hat C$ component which contains a term proportional to
the second derivative of $T_{ab\,ij}$. Also in view of later
applications we consider this term in more detail and note that 
the bosonic contribution to the second derivative of $T_{ab}^{ij}$
takes the form
\beq
D_\m D_c T_{ab}^{ij} = {\cal D}_\m{\cal D}_c T_{ab}^{ij} + f_{\m c}\,
T_{ab}^{ij} - 2f_{\m [a}\, T_{b]c}^{ij} + 2 f_{\m}^{\,d} \,\eta_{c[a} 
\,T_{b]d}^{ij} \,.
\eeq
Consequently 
\beq
D_\m D^a T_{ab}^{ij} = {\cal D}_\m{\cal D}^a T_{ab}^{ij} - f_{\m}^{\;a}
\, T_{ab}^{ij} \,. 
\eeq
With this result we consider the relevant term in $\hat C$,  
\bea
T^{ab\,ij} \,D_a D^c T_{cb\,ij}  = T^{ab\,ij} \, {\cal D}_a{\cal D}^c
T_{cb\,ij} - f_{a}^{\;c}\, T^{ab\,ij} \,T_{cb\,ij} \,, 
\label{hatC}
\eea
where we note in passing that, in the first term on the right-hand
side, we can symmetrize the derivatives as the antisymmetric part
vanishes due to the (anti-)selfduality  of the $T$-fields. By using
the equations found above, we can work out the double derivative on the
$T$-field, and verify whether it vanishes against the second term
proportional to $f_\m^{\,a}$. 

Rather than determining  $f_\m^{\,a}$ in this way, we 
continue and consider the supersymmetry variation of the
supercovariant derivatives
of fermion fields. First we make the observation that the derivatives
of $S$-invariant combinations of fields, whose $Q$-supersymmetric
variations were already required to vanish in the bosonic background,
will still vanish. But we can also compare the variation of the
supercovariant derivative of a fermion field to the variation of a
fermion field without derivatives. Consider for
example the $Q$-variation of the following $S$-invariant expression
\beq
D_\m \zeta^{\scriptscriptstyle\rm H}_i 
+ (-\ft14 \chi^{-1}{\cal D}\!\llap/\,\chi\,\d_i^j  + \ft12
k\!\llap/\,_{ik} \varepsilon^{kj})\,\g_\m \zeta^{\scriptscriptstyle\rm
H}_j\,.
\label{Dzeta}
\eeq
The derivative of another fermion field can now be written as the
derivative of an $S$-invariant linear combination of that fermion field with a
bosonic expression times $\zeta^{\scriptscriptstyle\rm H}_i$, which is 
one of the previously considered linear combinations whose vanishing
variation in the supersymmetric background has already been ensured, a
term proportional to \eqn{Dzeta} and terms proportional to
$\zeta^{\scriptscriptstyle\rm H}_i$ without a derivative. Therefore, once
we have imposed that the variation of \eqn{Dzeta} vanishes, then 
the variation of the derivative of every other fermion field is
guaranteed to vanish against some bosonic term times the variation of
$\zeta^{\scriptscriptstyle\rm H}_i$. Consequently variations of such
linear combinations can be ignored and our only task is to require
that the variation of \eqn{Dzeta} vanishes. Note that the above
argument can be extended to variations of multiple derivatives as
well, which therefore can also be ignored. 

Imposing  the condition that the $Q$-supersymmetry variation of
\eqn{Dzeta} vanishes, we find that most terms vanish already by virtue
of previous results and we are left with just one more equation,
\beq
D_\m \Big(\chi^{-1} D^a \chi\Big)  = \ft12 \Big({\chi}^{-1}\,D_\m
{\chi}\Big) \Big({\chi}^{-1} D^a {\chi}\Big) - \ft14 e_{\m}^{\, a}
\,\Big({\chi}^{-1}\,D_c {\chi}\Big)^2  \,.  
\eeq 
Note that we have superconformal derivatives here which involve the
gauge field $f_\m{}^{\!a}$ associated with conformal boosts. Upon using
the previous results \eqn{chiK}, \eqn{XK} and \eqn{AK}, all equations
coincide. Hence we are left with the following equation for $f_\m^{\,a}$,
\beq
f_\m{}^{\!a} =  -\ft12 {\cal D}_\m \Big( {\rm e}^{\cal K}\,{\cal D}^a
{\rm e}^{-{\cal K}}\Big) + \ft14 \Big({\rm e}^{\cal K}\,{\cal D}_\m
{\rm e}^{-{\cal K}}\Big) \Big( {\rm e}^{\cal K}\,{\cal D}^a {\rm
e}^{-{\cal K}}\Big)  - \ft18 e_{\m}^{\,a} \,\Big({\rm e}^{\cal K}\,
{\cal D}_c {\rm e}^{-{\cal K}}\Big)^2 \,,
\label{f-eq}
\eeq
which is $K$-invariant.
With this result we can verify that the term \eqn{hatC} vanishes as
well, so that we establish that the $\hat C$ component of the Weyl
multiplet vanishes. The above equation \eqn{f-eq} can be rewritten as 
\beq
R(\omega,e)_{\m}{}^a - \ft16 R(\omega,e) \,
e_\m^{\,a} = - \ft1{8} T_{\m b}^{ij}\, T^{ab}_{ij} + {\cal D}_\m 
{\cal D}^a {\cal K}
+\ft12 {\cal D}_\m {\cal K}\,
{\cal D}^a{\cal K} - \ft14 e_{\m}^{\,a} \,({\cal D}_c {\cal K})^2
\,. 
\eeq

So far the analysis is valid for any chiral background field. For the
rest of this section we assume that the chiral multiplet is given by
\eqn{background-def} so that at this point we have identified all
supersymmetric configurations in the presence of $R^2$-terms. The results
obtained so far are in a manifestly conformally covariant form. We  
can now impose gauge choices and set
$b_\m=0$ (because of the $K$-invariance the conditions found above are
in fact independent of $b_\m$) and $\exp[{\cal K}]$ equal to a
constant. (Alternative we may use $\exp[{\cal K}]$ as a 
compensator to make all quantities invariant under scale
transformations, at which point the field $b_\m$ will drop out.)  
The values of $\exp[-{\cal K}]$ and $\chi$ are related. With the choice   
that we made for the action we find that $\chi = -2 \exp[-{\cal K}]$ as
a result of the field equation for the field $D$. For future
reference, we give both the field equations for the field $D$ and for
the U(1) gauge field $A_\m$,  
\bea
3\,{\rm e}^{-\cal K} + \ft12 \chi &=& - 192 i\, D(F_A-\bar F_A)  
\nonumber\\
&&+ 4 i\Big\{ ( \varepsilon_{ij}\,T_{cd}^{ij})^{-2}\,
\varepsilon_{kl} T^{abkl} \, (F_I\,F^{-I}_{ab} - X^I \,G^-_{abI}) -
\mbox{h.c.} \Big\} \,, \label{D-eq} \\
 {\rm e}^{- {\cal K}} {\cal A}_a &=&  128 i \, {\cal D}^b 
\Big( F_A \, R(A)^-_{ab} 
- \mbox{h.c.} \Big) 
 - 8\, {\cal D}_c (F_A + {\bar F}_A) \, 
T_{ij ab}\,   T^{ijcb}  \nonumber\\
&&+ 8 (F_A - {\bar F}_A) \, \Big(T_{ab}^{ij}
 {\cal D}_c T_{ij}^{cb}
- T_{ijab}\, {\cal D}_c T^{cb ij} \Big) \nonumber\\
&&- 8 \,{\cal D}^b \Big\{  ( \varepsilon_{ij}\,T_{de}^{ij})^{-2}\,
\varepsilon_{kl} T^{kl\,c}{}_{[a} \, (F_I\,F^{-I}_{b]c} - X^I
\,G^-_{b]cI}) + \mbox{h.c.} \Big\}\,.
\label{A-eq}
\eea
Observe that these field equations can only be found from the action,
and cannot be obtained from requiring that the supersymmetry
variations vanish, because the action consists of a 
linear combination of two actions that are separately invariant,
corresponding to the vector multiplets and the hypermultiplets,
respectively (we point out that the hypermultiplets contribute only
fermionic terms to \eqn{A-eq}, which have been suppressed above). The
coefficient of the Ricci scalar in the action is now  
equal to $-(16\pi)^{-1} \exp[-{\cal K}]$, so that Newton's constant
equals $G_{\rm N} = \exp[{\cal K}]$, assuming a conventionally
normalized flat metric. Furthermore we can put the gauge fields $A_\m$
and  ${\cal V}_\m{}^{\!i}{}_j$ to zero, because their field strengths
vanish.

The most general $N=2$ supersymmetric background can now
be characterized as follows. First of all the spacetime has zero Weyl
tensor and is thus conformally flat. Its Ricci tensor is given by 
\beq
R_{\m\n} =- \ft18 T^{ij}_{\m\rho}\,T_{ij\n}{}^\rho \,,
\label{einstein}
\eeq
where  $T_{ij\m\n}$ ($T^{ij}_{\m\n}$) is a covariantly constant
(anti-)selfdual tensor. Furthermore we have a
number of constants $X^I$. The electric/magnetic field strengths
are also covariantly constant and given by  
\beq 
F_{\m\n}^{-I} =  \ft14 \varepsilon_{kl} T^{kl}_{\m\n} \,\bar X^I\,, 
\qquad
G_{\m\n I}^- =   \ft14 \varepsilon_{kl} T^{kl}_{\m\n} \,\bar F_I\,.
\label{FGT-eq}
\eeq

By using relations for products of (anti-)selfdual tensors one can
verify that the integrability condition that follows from the
covariant  constancy of the tensor fields $T_{\m\n}^{ij}$, is
identically satisfied. In order to investigate explicit solutions one
chooses coordinates such that the metric reads 
\beq
g_{\m\n} = {\rm e}^{2f(x)+{\cal K}} \,\eta_{\m\n} \,,
\label{confm}
\eeq
with $\eta_{\m\n}$ the flat Minkowski metric (normalized in the
standard way). We included the factor $\exp[{\cal K}]$, which we
adjusted to a constant, so that the
function $f$ is independent of the scale. 
To have a vanishing Ricci scalar the function $\exp [f]$ must be 
harmonic,
\beq  
\eta^{\m\n}\pa_\m \pa_\n  \,{\rm e}^f =0\,.
\eeq
The remaining conditions are (here we raise and lower indices with the 
flat metric) 
\bea
R_{\m\n} &= & 2\pa_\m\pa_\n f  - 2 \pa_\m f \,\pa_\n f  + 
\eta_{\m\n} \, (\pa_\rho f)^2  = - \ft18
T^{ij}_{\m\rho}\,T_{ij\n}{}^\rho\, {\rm e}^{-2f- {\cal K}}  
\,, \nonumber \\
\pa_\m T_{\n\rho}^{ij} &=& 2 \pa_\m f\,T^{ij}_{\n\rho} - 2 \pa_{[\n} f
\,T^{ij}_{\rho] \m} + 2 \eta_{\m[\n} \,T^{ij}_{\rho]\sigma}
\,\pa^\s f\,.
\eea
As a result of the second condition we derive
\beq
\pa_{[\m} T^{ij}_{\n\rho]} = \pa^\m T^{ij} _{\m\n} =0\,,
\eeq
so that $T_{\m\n}^{ij}$ is a harmonic tensor.
         
We are interested in time-independent solutions so that we  
assume that $f$ is independent of the time $t$. 
In that case we can express the tensor field in terms of a complex
potential $\Phi$. Denoting  spatial world indices by $\hat a,\hat
b,\hat c$, we may write  
\beq 
\varepsilon_{ij}T_{\hat a\hat b}^{ij} = \varepsilon_{\hat a\hat b\hat
c} \,\pa_{\hat c} \Phi \,,\qquad 
\varepsilon_{ij}T_{t\hat a}^{ij} = i \pa_{\hat a} \Phi\,,
\eeq
where $\Phi$ is a complex harmonic function. The equations are now solved
for by 
\beq
\Phi = 4 \,z\,{\rm e}^{f +{\cal K}/2}  \,, 
\eeq
with $z$ a constant phase factor, and $f$ satisfying
\beq
{\rm e}^f\, \pa_{\hat a}\pa_{\hat b} {\rm e}^f = 3 \,\pa_{\hat a} {\rm
e}^f\,\pa_{\hat b}{\rm e}^f - \d_{\hat a\hat b}  \,(\pa_{\hat c} {\rm
e}^f)^2 \,. \label{diff-eq}  
\eeq
This system of differential equations can be integrated. Its solution
is unique (up to translations) and is given by $\exp[f(r)]=
c/r$, where $c$ is a real constant. This leads to a Bertotti-Robinson
spacetime, the geometry of which describes the near-horizon limit of
an extremal black hole with horizon at $r=0$. Thus there exist no
fully supersymmetric multi-centered solutions, which 
is not suprising in view of the fact that the differential equations
\eqn{diff-eq} are nonlinear in $\exp[f]$. The field $\hat A$ is now
equal to 
\beq
\hat A = (\varepsilon_{ij} T^{ij}_{ab})^2 = {64\;{\rm e}^{-{\cal
K}}\over {\bar z}^2\,{\rm e}^{2f(r)}} \; (\pa_{\hat a} f)^2\,.
\eeq
{From} evaluating \eqn{FGT-eq} it follows
that the electric and magnetic charges are equal to 
\beq\label{eq:charges4}
p^I =  c\, {\rm e}^{{\cal K}/2}\,[\bar z\, X^I + z\, \bar X^I]
\,, \qquad  
q_I =  c\,{\rm e}^{{\cal K}/2}\,[\bar z\, F_I  +z\, \bar F_I] \,. 
\eeq
With this result we consider the so-called BPS mass, which takes the
form 
\beq 
\label{holo-bps-mass}
Z= {\rm e}^{{\cal K}/2} (p^I\,F_I- q_I\, X^I) =  - i z \,c \,,
\eeq
so that we obtain the equations (sometimes called stabilization
equations) \cite{FKS, BCDWKLM},
\beq
\bar Z\,\pmatrix{X^I\cr F_I\cr} - Z \,\pmatrix{\bar X^I\cr \bar
F_I\cr} = i \,{\rm e}^{-{\cal K}/2} \,\pmatrix{p^I\cr q_I\cr}\;.
\eeq
Observe that this result is covariant with respect to
electric/magnetic duality. 

Finally we note that the area in Planck units equals 
\beq
{{\rm Area}\over G_{\rm N}} = 4\pi\,c^2  = 4\pi\, \vert Z\vert^2\,. 
\eeq
This does not determine the black hole entropy, because the
Bekenstein-Hawking area law is not applicable for these 
black holes \cite{Wald}. 
After including an appropriate correction one obtains instead \cite{CDM}
\beq
{\cal S} = \pi\Big[ \vert Z\vert^2 - 256 \,{\rm Im}
[F_A(X^I,\hat A)]\Big]\,,
\label{entropy}
\eeq
where $\hat A = - 64\, \bar Z^{-2} \,{\rm e}^{-{\cal K}}$. 

In section \ref{sec:N=1} we will be using another coordinate frame with line
element given by 
\begin{equation}
  \label{eq:lineel5}
  {\rm d}s^2 = - {\rm e}^{2g} \, {\rm d} t^2 + {\rm e}^{-2g}\, {\rm d} \vec{x}^2\,.
\end{equation}
The conformal coordinates of this section are related to those of the above frame by
\begin{equation}\label{eq:frame4to5}
  t \longrightarrow {d \over c^2 \,{\rm e}^{{\cal K}} } \,  t\,, \quad \vec{x}
  \longrightarrow d\, {\vec{x}\over |\vec{x}|^2}\,, 
\end{equation}
where $d$ is some real constant. 
The function ${\rm e}^{-2g}$ in \eqn{eq:lineel5} corresponding to the line
element \eqn{confm} is equal to 
\begin{equation}
   {\rm e}^{-2g} = {c^2\,{\rm e}^{{\cal K}}
  \over |\vec{x}|^2}\,.
\end{equation}
For later reference
let us give the field strengths \eqn{FGT-eq} in the frame \eqn{eq:lineel5},
\begin{equation}\label{eq:fs4}
  F^{-\,I}_{tm} =  i z \bar X^I \, {\rm e}^{g}\, {x^m \over |\vec{x}|^2 }\,,
  \qquad   G^{-}_{tm\,I} =  i z \bar F_I \, {\rm e}^{g}\, {x^m \over |\vec{x}|^2 }\,.
\end{equation}
Here $(t,m)$ denote world indices in the frame \eqn{eq:lineel5}. For these
expressions Maxwell's equations \eqn{eq:normcharges} are satisfied with the
charges defined in \eqn{eq:charges4}. Observe that, when calculating Maxwell's
equations directly from \eqn{FGT-eq} in the frame \eqn{confm}, one encounters a
different sign as compared to \eqn{eq:charges4}.  This is related to
the fact that a charge located at the origin in the frame \eqn{eq:lineel5}
corresponds to a charge at infinity in the conformal coordinates used in this
section. When evaluating Maxwell's equations in the latter coordinates one is
considering the corresponding mirror charge placed at the origin. This
explains the apparent sign discrepancy.
 
\section{$N=1$ supersymmetric field configurations}\label{sec:N=1}
\setcounter{equation}{0}
A general analysis of the conditions for residual $N=1$
supersymmetry is extremely cumbersome. Therefore we base ourselves on
a given class of embeddings of the residual supersymmetry by imposing
the following condition on the supersymmetry parameters, 
\begin{equation}
h\, \epsilon_i =  \ve_{ij} \,\gamma_0 \,\epsilon^j\,,
\label{N=1-condition}
\end{equation}
where $h$ is some unknown phase factor which is in general not
constant, and which transforms under U(1) with the same weight as the
fields $X^I$. At the moment we proceed without imposing gauge
choices. Therefore the choice of $\gamma_0$ is somewhat arbitrary,
because it can be changed into any other gamma matrix by means of a
local Lorentz transformation. However, we will eventually impose a
gauge condition on the vierbein field, which restricts the local Lorentz
transformations to the three-dimensional rotations\footnote{
  In view of this, Lorentz covariant derivatives should be applied
  with caution, as the various equations we are about to derive are
  not Lorentz covariant. 
}. 
It is clear that
\eqn{N=1-condition} is then consistent with spatial rotations and SU(2)
transformations, although we will not require the solutions
to be invariant under these symmetries. An  embedding condition such
as \eqn{N=1-condition} was also used in the analysis presented in 
\cite{Sabra,Moore} of $N=2$ theories without $R^2$-interactions.

Subject to this embedding we can now evaluate the conditions for $N=1$
supersymmetry by following essentially the same steps as in the previous
section. We start by considering the variations of the vector multiplet
fermions and of the spinors $\zeta_i^{\scriptscriptstyle \rm V}$ and
$\zeta_i^{\scriptscriptstyle \rm H}$. They lead to the equations 
\beq
\hat{B}_{ij}= k_{a\,ij}= 0\,, \eeq and \bea {\cal A}_0 = 0\,,\qquad&&
{\cal A}_p =  {\rm Re}[{h}\,  {\cal F}_{0p}^-]\,, \nonumber \\
{\cal D}_0 ( \chi {\rm e}^{{\cal K}}) = 0\,, \qquad&& {\cal D}_p ( \chi {\rm
  e}^{{\cal K}}) = 2 \,\chi {\rm e}^{{\cal K}} \,{\rm Im}[{h}\, {\cal
  F}_{0p}^-] \,,
\label{A-F-eqs}
\eea
where the indices $(0,p)$ with $p = 1,2,3$ refer to the tangent
space. With this result we find that the variation of   
$\zeta_i^{\scriptscriptstyle \rm V}$ simplifies considerably and
reduces to
\beq
\d\zeta_i^{\scriptscriptstyle \rm V}  =   \chi^{-1}{\cal D}\!\llap/\,
\chi \,\e_i  +2  \,\eta_i\,. \label{compS}
\eeq
For the hypermultiplets we find the same condition as for full
supersymmetry,
\beq
{\cal D}_a (\chi^{-1/2} A_i{}^\a) = 0\,.
\eeq

Returning to the vector multiplet spinors, we then establish the
relations
\beq
{\cal D}_0( \chi^{-1/2}\, X^{I}) =  {\cal D}_0(\chi^{-1/2}\,
F_I ) = 0\,, 
\eeq
and 
\begin{eqnarray}\label{Bogo-eq}
{\cal D}_p ( \chi^{-1/2} \,X^{I}) &=&  - h\, \chi^{-1/2}
( F_{0p}^{-I} -\ft14 \varepsilon_{kl} T^{kl}_{0p} \,\bar X^I)\,,
 \nonumber\\
{\cal D}_p (\chi^{-1/2} \,F_I) &=& - h\, \chi^{-1/2}( G^-_{0pI}-
 \ft14 \varepsilon_{kl} T^{kl}_{0p} \,\bar F_I)\,.  
\eea
These last two equations transform covariantly with respect to
electric/magnetic duality. Taking their symplectically invariant
product with $(\bar X^I,\bar F_I)$ leads to the previous equations
\eqn{A-F-eqs}. 

Subsequently we consider the variations of the spinor $\chi^i$, which
lead to 
\bea
R({\cal V}) _{ab}{}^{\!i}{}_j &=& 0\,,\nonumber \\
{\cal D}_c(\chi^{1/2} \,T^{ijc0}\,\ve_{ij}) &=& - 6 h\,\chi^{1/2}
\,D\,,\nonumber \\
{\cal D}_c(\chi^{1/2} \,T^{ijcp}\,\ve_{ij}) &=& 8 i h\,\chi^{1/2}
\,R(A)^{-0p}\,.
\label{chi-vanishing}
\eea
Note that the first equation is consistent with the fact that $\hat
B_{ij}$ vanishes (c.f. \eqn{background-def}). In view of the fact that
the SU(2) field strengths vanish, we will set the SU(2) connections to
zero in what follows. 

The variations for the field strength $R(Q)^i_{ab}$ lead to 
\bea
&&{\cal D}_0 T_{ab}^{ij} -\ft12 \chi^{-1}{\cal D}_d \chi\,\Big( \d^d_0\,
T^{ij}_{ab} 
- 2\d^d_{[a} \,T^{ij}_{b]0} + 2 \eta_{0[a}\,T^{ij\,d}_{\,b]}\Big) =0  \,,  
\nonumber \\    
&&{\cal D}_p T^{ij}_{ab}  -\ft12 \chi^{-1}{\cal D}_d \chi \,\Big( \d^d_p\,
T^{ij}_{ab} - 2\d^d_{[a} \,T^{ij}_{b]p} + 2
\eta_{p[a}\,T^{ij}_{\,b]}{}^d\Big)  
= 4 h \,\ve^{ij} \, {\cal R}(M)^{-}_{a b \,0 p} \,.
\label{RQ-vanishing}
\eea

Finally we consider the variation of derivatives of fermion
fields. The arguments presented below \eqn{Dzeta} about the fact that
there is no need to consider more than one of these variations, apply
also to residual supersymmetry. Hence we consider the
$Q$-supersymmetry variation of \eqn{Dzeta}, making use of the
previously obtained results. This yields the following equation, 
\bea
&&D_\m (\chi^{-1} D^a\chi) +\ft14 (\chi^{-1} D_c\chi)^2
\,e_\m{}^a -\ft12 (\chi^{-1}D_\m\chi) (\chi^{-1}D^a\chi)= 
\nonumber\\ 
&&-\ft32 D(e_\m{}^a  - 2 e_{\m 0}\,\eta^{a0}) - 2i[R(A)^+ -R(A)^-]_\m{}^a
-4i \,R(A)^-_{\m 0} \,\eta^{a0}  \,. 
\label{DDchi}
\eea
All terms in this equation are real, with the exception of the last
term, from which it follows that $R(A)^\pm_{a0}$ must be purely
imaginary, so that
\beq
R(A)_{a0} = \tilde R(A)_{pq} =  0\,.
\label{RA-vanishing} 
\eeq
Just as before, \eqn{DDchi} fixes the value of the gauge field
$f_\m{}^a$, which takes the ($K$-invariant) form
\bea
f_\m{}^a &=&-\ft12 {\cal D}_\m (\chi^{-1}{\cal D}^a\chi) -\ft18
(\chi^{-1}{\cal D}_c\chi)^2 
\,e_\m{}^a +\ft14 (\chi^{-1}{\cal D}_\m\chi) (\chi^{-1}{\cal
D}^a\chi) \nonumber\\ 
&&-\ft34 D(e_\m{}^a - 2 e_{\m 0}\,\eta^{a0}) -i [R(A)^+ -R(A)^-]_\m{}^a
-2i\,R(A)^-_{\m 0} \,\eta^{a0}  \,.
\label{f-result}
\eea
Comparing with \eqn{f-def} yields
\bea
&&R(\omega,e)_{\m}{}^a - \ft16 R(\omega,e) \,
e_\m^{\,a} = \nonumber\\
&&\hspace{10mm} 
- {\cal D}_\m (\chi^{-1}{\cal D}^a\chi) -\ft14
(\chi^{-1}{\cal D}_c\chi)^2 
\,e_\m{}^a +\ft12 (\chi^{-1}{\cal D}_\m\chi) (\chi^{-1}{\cal
D}^a\chi)  \nonumber\\
&&\hspace{10mm}  - \ft1{8} T_{\m b}^{ij}\, T^{ab}_{ij} 
- D\, (e_\m{}^a - 3 e_{\m 0}\,\eta^{a0}) +i[\tilde R(A)_{\m 0}\,\eta^{a0} -
\tilde R(A)_0{}^a\,e_\m{}^0]  \,.
\label{Ricci}
\eea

Let us briefly return to \eqn{chi-vanishing} and \eqn{RQ-vanishing}
and explore the consequences of \eqn{RA-vanishing}. 
The first equation of \eqn{RQ-vanishing}  yields
\beq
{\cal D}_0 T^{ij\,0p}- \ft12 \chi^{-1}{\cal D}_0\chi \,T^{ij\,0p} + \ft12
\chi^{-1}{\cal D}_q \chi \,T^{ij\,qp}   = 0\,. \label{D0T}
\eeq
Making use of this, the last equation \eqn{chi-vanishing} leads to
\beq
{\cal D}_q T^{ij\,qp}+ \chi^{-1}{\cal D}_0\chi \,T^{ij\,0p}  =
-2ih\,\varepsilon^{ij} \,\tilde R(A)^{0p} \,, 
\eeq
which can be rewritten as
\beq
{\cal D}_{[p}T^{ij}_{q]0}\,\varepsilon_{ij} = 2i R(A)_{pq} h - 
\ft12 \chi^{-1}{\cal D}_0\chi\, T_{pq}^{ij}\,\varepsilon_{ij}\,.
\label{curl-T}
\eeq

Observe that so far we have not imposed any gauge conditions. In order
to proceed we will now choose a gauge condition that eliminates the
freedom of making (local) scale transformations and conformal
boosts. This gauge condition amounts to choosing $b_\m=0$ and $\chi$
constant. Therefore the covariant derivative ${\cal D}_a$ contains
only the spin connection fields and the U(1) connection,
when appropriate. 

In this gauge, \eqn{curl-T} and the second equation of
\eqn{chi-vanishing} read, 
\beq
\bar h\, {\cal D}_{[p}T^{ij}_{q]0}\,\varepsilon_{ij} = 2i
R(A)_{pq}\,,\qquad 
\bar h\,{\cal D}^p T^{ij}_{p0} \,\ve_{ij} = 6 \,D\,.
\label{RA-D}
\eeq
Furthermore we establish from \eqn{Ricci} that
\beq
R(\omega,e) =  - 3D \,.
\label{rd}
\eeq
Then, from the second equation of \eqn{RQ-vanishing}, one derives the
following expressions for the components of the curvature tensor ${\cal
R}(M)_{ab\,cd}$,   
\bea
{\cal R}(M)_{pq\, 0r}&=&  \ft18 i \varepsilon_{pq}{}^s \, \bar h
\,{\cal D}_r T^{ij}_{s0}\,\varepsilon_{ij} + \mbox{ h.c.}\,,   \nonumber\\
{\cal R}(M)_{0r\, pq}&=& \ft18 i \varepsilon_{pq}{}^s \, \bar h
\,{\cal D}_s T^{ij}_{r0}\,\varepsilon_{ij} + \mbox{ h.c.}\,, 
 \nonumber\\
{\cal R}(M)_{0p\, 0q}&=&  - \ft18  \bar h
\, {\cal D}_q T^{ij}_{p0}\,\varepsilon_{ij} + \mbox{ h.c.}\,, \nonumber\\
{\cal R}(M)_{pq\,rs}&=&  \ft18  \varepsilon_{rs}{}^v
\varepsilon_{pq}{}^u \, \bar h
\,{\cal D}_v T^{ij}_{u0} \,\varepsilon_{ij} + \mbox{ h.c.}\,. 
\eea
These expressions satisfy all the constraints \eqn{RM-constraints}
listed in appendix \ref{b}, provided 
one makes use of the relations for $R(A)$ and $D$ (cf. \ref{RA-D}).  
Using \eqn{f-result} and the definition of ${\cal R}(M)$ allows us to
find expressions for the components of the Riemann tensor. Making use
of \eqn{RA-D} we find 
\bea
R(\omega)_{pq\, 0r}&=& R(\omega)_{0r\, pq}\nonumber\\
&=&  \ft18  \varepsilon_{pq}{}^s \, \Big[ i(
\bar h \,{\cal D}_r T^{ij}_{s0}\,\varepsilon_{ij} - \ft12 
T^{ij}_{r0} \,T_{ij\,s0}) + \mbox{ h.c.}\Big]  \,,   \nonumber\\
R(\omega)_{0p\, 0q}&=&  R(\omega)_{0q\, 0p} \nonumber\\ 
&=&
- \ft18 \Big[ (  \bar h
\, {\cal D}_q T^{ij}_{p0}\,\varepsilon_{ij} +\ft12
T^{ij}_{q0}\,T_{ij\,p0} ) + \mbox{ h.c.}\Big] \,, \nonumber\\
R(\omega)_{pq\,rs}&=&  -\ft12  \d_{[r[p} \Big[ \bar h
\,{\cal D}_{q]} T^{ij}_{s]0} \,\varepsilon_{ij} 
+ \mbox{ h.c.} \Big] \nonumber\\
&& + \ft14  \d_{[r[p} \Big[ T^{ij} _{q]0} \,
T_{ijs]0} + T_{ijq]0} \,T^{ij} _{s]0} -  \d_{q]s]} \,T^{ij
\,v}{}_{0}\,T_{ij\,v0}  \Big] \,. \label{rc}
\eea
Here we observe that, owing to \eqn{RA-D}, this result satisfies all
the algebraic properties of a Riemann tensor, such as cyclicity and
pair exchange.   
We also note that, by virtue of \eqn{RA-D},  (\ref{rc})
gives rise to (\ref{Ricci}) upon contraction.

At this point we adopt a gauge condition for local Lorentz
invariance. We remind the reader that the supersymmetry embedding
condition \eqn{N=1-condition} is obviously inconsistent with local
Lorentz invariance and presupposes that we would eventually impose 
such a gauge condition. Therefore we bring the vierbein
field in block-triangular form by imposing $e_t{}^p= 0$, thereby
leaving the SO(3) tangent-space rotations unaffected. Denoting world
indices by $(t,m)$, with $m=1,2,3$, we parametrize the vierbein as follows, 
\beq
e_\m{}^0 {\rm d}x^\m = {\rm e}^g [\,{\rm d}t + \sigma_m \,{\rm
d}x^m\,]\,,\qquad  e_\m{}^p {\rm d}x^\m = 
{\rm e}^{-g} \,\hat e_m{}^p\,{\rm d}x^m \,,
\eeq
where $\hat e_m{}^p$ is the rescaled dreibein of the three-dimensional
space. The corresponding inverse vierbein components are then given by 
\beq
{}e_0{}^t = {\rm e}^{-g} \,,\quad e_0{}^m=0\,,\quad e_p{}^t = -\sigma_p
\,{\rm e}^g \,, \quad e_p{}^m = {\rm e}^g\, \hat e_p{}^m \,,
\eeq
where, on the right-hand side, spatial tangent-space and world indices
are converted by means of the dreibein fields $\hat e_m{}^p$ and its
inverse. 

Now we concentrate on stationary spacetimes, so that we can adopt 
coordinates such that the vierbein components are independent of the
time coordinate $t$. In that case the spin connection fields take the
following form, 
\beq
\omega_{l\,pq}= {\rm e}^g [\, \hat \omega_{l\,pq} +2\d_{l[p}\,\nabla_{q]}
g\,] \,, \quad
\omega_{0\, pq}= \omega_{q\,p0}= -\ft12 {\rm e}^{3g} \,\varepsilon_{pql}
\,R(\sigma)^l  \,, \quad
\omega_{0\,0p}= {\rm e}^g\,\nabla_pg  \,,
\eeq
where $\hat\omega_m{}^{pq}$ is the spin-connection field associated
with the dreibein fields $\hat e$ in the standard way.
We used the definition  
\beq
R(\sigma)^l = \varepsilon^{lpq} \, \nabla_{p}\sigma_{q} \,.
\eeq
Observe that $\nabla^pR(\sigma)_p=0$. The covariant derivatives
$\nabla_m$ refer to the three-dimensional space only. Hence they
contain the three-dimensional spin connection $\hat\omega_m{}^{pq}$. 
  
The corresponding curvature components take the following form (where
we consistently use three-dimensional notation on the right-hand side), 
\bea
R(\omega)_{pq\,0r} &=& \ft12 \varepsilon_{pq}{}^s \, {\rm e}^{4g} \Big[
\nabla_r  R(\sigma)_s + 5 \,R(\sigma)_s\,\nabla_rg +
R(\sigma)_r\,\nabla_sg - 2\d_{sr} \, 
R(\sigma)^u\,\nabla_u g\Big] \,, \nonumber \\
R(\omega)_{0p\,0q} &=& -{\rm e}^{2g} \Big[\nabla_p\nabla_q g + 3\,
\nabla_pg\,\nabla_qg - \d_{pq}\,(\nabla_rg)^2\Big] 
 + \ft14 {\rm
e}^{6g} \Big[R(\sigma)_p\,R(\sigma)_q - \d_{pq} R(\sigma)^2\Big]
\,, \nonumber \\ 
R(\omega)_{pq\,rs} &=& {\rm e}^{2g} \, \hat R_{pq\,rs} 
- 4\,{\rm e}^{2g} \,\d_{[p[r}\Big[ \nabla_{s]}\nabla_{q]}g +
\nabla_{s]}g\,\nabla_{q]} g - \ft12\d_{s]q]} \, (\nabla_ug)^2 \Big]
\nonumber \\
&&
+3 \,{\rm e}^{6g} \, \d_{[p[r} 
\Big[R(\sigma)_{s]}\,R(\sigma)_{q]}-\ft12 \d_{s]q]} \,R(\sigma)^2
\Big]\,.   
\label{rome}
\eea

However, for stationary solutions also other quantities than those
that encode the spacetime should be time-independent. Hence we infer 
that $\bar h X^I$, $\bar h F_I$, and $\bar h T_{p0}^{ij}$ are
time-independent while $(\pa _t +i A_t)h=0$. 

Until now we have restricted our attention to quantities that are
supercovariant with respect to full $N=2$ supersymmetry. However, when
considering residual supersymmetry, certain linear combinations of the
gravitini will still transform covariantly. To see how this works, let us
record the gravitini transformation rules in the restricted
background. Here we make use of \eqn{compS} to argue that there is no
need for including compensating $S$-supersymmetry transformations. The
result takes the form
\bea
\d\psi_t^i&=& 2 \pa_t \e^i +i A_t \,\e^i + {\rm e}^{2g} \Big[ T_p -
\nabla_p g + \ft12 i {\rm e}^{2g} \,R(\sigma)_p  \Big] \,\gamma^p
\g_0\e^i \,, \nonumber \\ 
\d\psi_m^i&=& 2\nabla_m\e^i - ( T_m -iA_m ) \e^i
\nonumber \\
&& - i \hat e_m^{\;p}\, \varepsilon_{p}{}^{\!qr} \Big[ T_r -
\nabla_r g + \ft12 i {\rm e}^{2g} 
\,R(\sigma)_r  \Big] \,\gamma_q \g_0\e^i   \nonumber \\ 
&&+ \s_m\, {\rm e}^{2g} \Big[T_p - \nabla_p g + \ft12 i {\rm
e}^{2g} \,R(\sigma)_p\Big] \,\gamma^p \g_0\e^i \,,  
\label{Killing-spinor}
\eea
where we have introduced a three-dimensional world vector
$T_m$, 
\beq
T_m \equiv \ft14  {\rm e}^{-g} \,\hat e_m{}^p \,\bar h T^{ij}_{p0}
\,\varepsilon_{ij}  \,. 
\eeq
Now we observe that the combinations $\psi_{\m i} - \bar
h\,\varepsilon_{ij} \gamma_0\psi_\m^j$ transform covariantly under the
residual supersymmetry. {From} the requirement that these covariant
variations vanish we deduce directly that
\beq
T_m= \nabla_m g - \ft12 i {\rm e}^{2g}\,   R(\sigma)_m \,, \qquad
\bar h\nabla_m h + i A_m  = -\ft12 i {\rm e}^{2g}\,   R(\sigma)_m \,.
\eeq
This leads to the following expressions for the gravitini variations, 
\beq
\d\psi_t^i= 2 \pa_t \e^i + i A_t \e^i  \,, \qquad
\d\psi_m^i= 2\nabla_m\e^i - (\nabla_m g + \bar h\nabla_mh) \e^i \,.
\label{gravitini} 
\eeq

With these results we return to the previous identities and verify
whether they are now satisfied. It is straightforward to see that this
is the case for \eqn{D0T}. For the other identities one needs the
covariant derivative $\bar h\,{\cal D}_p T_{q0}^{ij}$, which,  in
three-dimensional notation,  reads
\beq
\bar h\,{\cal D}_pT_{q0}^{ij}\varepsilon_{ij} = 4\,{\rm e}^{2g}
\,\Big[  \nabla_p  T_q + 2\,T_pT_q - \d_{pq} \,(T_r)^2 \Big] \,.
\eeq
It is now straightforward to prove \eqn{RA-D} with $D$ given
by
\beq
D= \ft23 {\rm e}^{2g} \Big[ \nabla_p^{\,2} g - (\nabla_p g)^2 +\ft14
{\rm e}^{4g} \,(R(\sigma)_p)^2\Big]\,.  \label{D-result}
\eeq
Furthermore, it
turns out that \eqn{rc} and \eqn{rome} agree, provided that the
curvature of the three-space is zero, 
\beq
\hat R_{mn\,pq}=0\,,
\eeq
so that the three-dimensional space is flat. Observe that this result
is consistent with the integrability condition corresponding to the
Killing spinor equations that one obtains when setting the gravitino
variations \eqn{gravitini} to zero. The only remaining
equations are now \eqn{Bogo-eq}, which express the abelian field
strengths in terms of the other fields,
\begin{eqnarray}\label{Bogo-eq2}
F_{0p}^{-I} &=& -{\rm e}^g \Big[ \nabla_p (\bar hX^I) + (\nabla_p
g)\,h\bar X^I -\ft12 i {\rm e}^{2g}R(\sigma)_p (\bar hX^I+h\bar
X^I)\Big]\,,  \nonumber\\
G^-_{0pI}&=& 
-{\rm e}^g \Big[ \nabla_p (\bar hF_I) + (\nabla_p
g)\,h\bar F_I -\ft12 i {\rm e}^{2g}R(\sigma)_p (\bar hF_I+h\bar
F_I)\Big]\,,
\eea
where on the right-hand side, we consistently use three-dimensional
tangent space indices.  
With these results we derive the following expressions,
\bea
{\cal D}^a F^{-I}_{ap} &=& i {\rm e}^g \,\varepsilon_p{}^{\!qr}
\nabla_qF^{-I}_{0r} \nonumber\\
&=& -\ft12 {\rm e}^g   \,\varepsilon_p{}^{\!qr} \nabla_q\Big[ {\rm
e}^{3g} R(\sigma)_r (\bar h X^I + h\bar X^I) \Big] \nonumber\\
&&- i {\rm e}^{2g} \,\varepsilon_p{}^{\!qr} \,\nabla_qg \,\nabla_r
(\bar h X^I - h\bar X^I) \,,\nonumber\\
{\cal D}^aF^{-I}_{a0} &=& {\rm e}^g \Big[\nabla^qF^{-I}_{q0} - 2
(\nabla^qg -\ft12 i{\rm e}^{2g}
R(\sigma)^q)\,F^{-I}_{q0}\Big]\nonumber\\
&=& {\rm e}^{2g} \Big[ \nabla_p^{\,2} (\bar h X^I) +
(\nabla_p^{\,2}g) \,h\bar X^I - (\nabla_p g)^2 \,h\bar X^I +
(\nabla_pg) \,\nabla ^p(h\bar X^I-\bar h X^I) \nonumber\\
&&\hspace{5mm}  
-\ft12i{\rm e}^{3g} R(\sigma)^p \,\nabla_p[ {\rm e}^{-g} (h\bar X^I -
\bar h X^I)] + \ft12 {\rm e}^{4g} (R(\sigma)_p)^2 (h\bar X^I + \bar h
X^I) \Big] \,,\;
\eea
and likewise for the electric/magnetic dual equations (i.e., replacing
$F^{-I}$ by $G_I^-$, etcetera). The imaginary parts of the above
expressions correspond to Maxwell's equations for the abelian vector
fields. Because the first expression is manifestly real, the
corresponding Maxwell equation (and its electric/magnetic dual) is
satisfied. The imaginary part of the 
second expression and its dual equation provide the remaining
Maxwell equations, which read
\bea
\nabla_p^{\,2} \Big[{\rm e}^{-g} (\bar h X^I - h\bar X^I)\Big] =0\,,
\nonumber\\ 
\nabla_p^{\,2} \Big[{\rm e}^{-g} (\bar h F_I - h\bar F_I)\Big] =0\,, 
\label{harmonic}
\eea
which shows that the functions in parentheses are harmonic. 
Furthermore we note the equations 
\bea
F_{0p}^{-I}+F_{0p}^{+I} &=& - \nabla_p\Big[{\rm e}^g  (\bar hX^I+
h\bar X^I) \Big]\,,  \nonumber\\
G^-_{0pI}+ G^+_{0pI}&=& 
- \nabla_p \Big[ {\rm e}^g(\bar hF_I + h\bar F_I)\Big] \,,
\eea
so that the functions under the derivative can be regarded as electric
and magnetic potentials. 

So far our analysis is valid for any chiral background. Now we
identify this background with \eqn{background-def} and note that the
field $\hat A$ can be written as  
\beq 
\hat A = - 64 {\rm e}^{2g} \,h^2 \, (T_p)^2\,.
\eeq
With this choice for the background we now evaluate the field
equations for the fields $D$ and $A_\m$, which were given in
\eqn{D-eq} and 
\eqn{A-eq}, respectively. 
Using  \eqn{Bogo-eq}, \eqn{D-result}  and the homogeneity properties
of $F(X, {\hat A})$, the first equation takes the form
\bea
{\rm e}^{-\cal K} + \ft12 \chi &=& -  128 i \,{\rm e}^{3g} \,\nabla^p
\Big[ {\rm e}^{-g}\, \nabla_pg \,(F_A-\bar F_A)\Big] - 32 i \,{\rm e}^{6g}
\,(R(\sigma)_p)^2 (F_A-\bar F_A) \nonumber \\
&& -64 \,{\rm e}^{4g} R(\sigma)_p\,\nabla^p  (F_A +\bar F_A ) \,.
\label{dk}
\eea
The second equation \eqn{A-eq} comprises four equations. The one with
$a=0$ turns out to be identically satisfied, by virtue of
of an intricate interplay of all the results that we obtained
above. This constitutes a very subtle check upon the correctness of
the results obtained so far. Using similar manipulations the equation
\eqn{A-eq} 
with $a=p$ can be written as
\begin{eqnarray}
&& ({\bar h} X^I - 
h {\bar X}^I) \stackrel{\leftrightarrow}\nabla_p ({\bar h} F_I - 
h {\bar F}_I) -\ft12 \chi \, {\rm e}^{2g} \, R(\sigma)_p = \nonumber\\
&&\hspace{5mm} 128 \, 
{\rm e}^{2g}\,  \nabla^q \Big[ 2 \nabla_{[p} g \,\nabla_{q]} (F_A +
{\bar F}_A)  + i 
\nabla_{[p}  \Big( {\rm e}^{2g} \, R(\sigma)_{q]}\,(F_A - {\bar F}_A)  
\Big) \Big] 
\,. \label{ak}
\end{eqnarray} 
To arrive at this concise expression requires an extensive usage of
many of the previously obtained results, and in particular of
(\ref{dk}). 

This concludes our analysis. The solutions can now be expressed in
terms of harmonic functions according to \eqn{harmonic}. The two
field equations \eqn{dk} and \eqn{ak} then determine the function $g$ and
$R(\sigma)_p$, from which all other quantities of interest
follow. We should point out that there are some equations of motion 
whose validity has not yet been verified. We claim that those are implied
by the residual supersymmetry of our solutions. For instance, for the
vector multiplets we have imposed the Maxwell equations. Therefore the
$N=1$ supersymmetry variation of the field equations of the vector multiplet
fermions can only lead to the field equations of the vector multiplet
scalars, which must thus be satisfied by supersymmetry. For the
hypermultiplets a similar argument holds. Indeed, the result \eqn{rd},
which is crucial for the validity of the field equation for the
hypermultiplet scalars, has already been established on the basis of
the previous analysis. The field equations for the fields of the Weyl
multiplet have been imposed, with the exception of those for the
vierbein field and the tensor field $T_{ij ab}$ (the field equations
for the SU(2) gauge fields are trivially satisfied because of the
SU(2) symmetry of our solutions). However, the field equations of the
gravitino fields and of the fermion doublet $\chi^i$ transform into
these two field equations, from which one may conclude that they are
also satisfied by supersymmetry.

\section{Discussion and conclusions}\label{sec:conclusions}
\setcounter{equation}{0}
In this paper we have characterized all stationary solutions with a
residual $N=1$ supersymmetry embedded according to
\eqn{N=1-condition}. In principle there may exist other solutions based on
inequivalent embeddings of $N=1$ supersymmetry. It should be
interesting to apply our approach to more general embeddings
of the residual supersymmetry. 

By imposing the conditions for residual
supersymmetry and a subset of the field equations we have obtained the
full class of these solutions, albeit not explicitly because the
equations depend on the holomorphic function $F(X,\hat A)$  that
characterizes both the vector multiplets and the $R^2$-interactions. A
gratifying feature of our results is that the  
presence of the $R^2$-interactions gives rise to relatively minor
complications, something that may seem rather surprising in view of
the complicated structure of the higher-derivative terms in the
action. There are two reasons for the fact that these complications
can remain so implicit in our analysis. The first is that
the higher-derivative interactions are nicely encoded in the
holomorphic function $F(X,\hat A)$. The second reason is that we have
consistently used quantities that transform covariantly under
electric/magnetic duality. Without this guidance there would be a
multitude of ways to express our results and perform the analysis. 

We have also shown that solutions with supersymmetry enhancement
exhibit fixed-point behavior of the moduli fields, simply because
the solutions with full $N=2$ supersymmetry are unique. This result is
relevant when calculating the horizon geometry of extremal black holes
since it explains why the black hole entropy depends only on the
electric and magnetic charges carried by the black hole.

Let us briefly summarize the solutions that we have found. 
Following \cite{BCDWKLM} we introduce the rescaled U(1) and Weyl
invariant variables,  
\bea
Y^I =  {\rm e}^{-g}  \, {\bar h} \, X^I \;,\qquad 
\Upsilon = {\rm e}^{-2 g}  \, {\bar h}^2 \, {\hat A} \,,
\label{yvar}
\eea 
so that, using the homogeneity of  $F(X,\hat A)$, we can write
$F(Y,\Upsilon) = \exp[-2g] \,\bar h^2 F(X,\hat A)$ and 
\begin{equation}
   \pmatrix{Y^I\cr\noalign{\vskip2mm} F_I(Y,\Upsilon)} = {\rm
e}^{-g}\,  \bar h 
\pmatrix{X^I\cr\noalign{\vskip2mm}
   F_I(X,{\hat A})}\,.
 \end{equation}
Observe that $F_A(X,\hat A) = F_\Upsilon(Y,\Upsilon)$. Henceforth we
will always use the rescaled variables. 
The rescaled background field $\Upsilon$ is given by
\beq
\Upsilon= - 64 \Big(\nabla_m g - \ft12 i {\rm e}^{2g}\,
R(\sigma)_m\Big)^2\,. 
\eeq
Furthermore from (\ref{kaehler}) and (\ref{yvar}) we infer that 
\bea\label{eq:line}
{\rm e}^{-2g} =i\,  {\rm e}^{\cal K}  
\,  \Big[{\bar Y}^I F_I (Y, \Upsilon)
- {\bar F}_I ({\bar Y}, {\bar \Upsilon}) {Y}^I\Big] \,.
\eea
According to \eqn{harmonic} we can express the imaginary part of 
$(Y^I,F_J)$ in terms of a symplectic array of $2(n+1)$ harmonic
functions $(H^I(\vec x),H_J(\vec x))$,  
\begin{equation}
    \pmatrix{Y^I-{\bar Y}^I\cr\noalign{\vskip2mm} F_I(Y,\Upsilon)-{\bar
    F}_I(\bar Y,\bar \Upsilon) } = i
    \pmatrix{H^I\cr\noalign{\vskip2mm}H_I}\,. \label{gen-stab-eqs} 
\end{equation}
These are the ``generalized stabilization equations'' which were
conjectured in \cite{Sabra} and \cite{BCDWLMS} for the case without
and with $R^2$-interactions respectively (a derivation for certain solutions
without $R^2$-terms appeared in \cite{Denef}). In principle these equations determine the
full spatial dependence of 
$Y^I$  in terms of the harmonic functions and the
background field $\Upsilon$. However, explicit 
solutions of the stabilization equations can only be obtained in a
small number of cases and usually one has to solve the equations by
iteration which is extremely cumbersome.  We will discuss a few examples
of explicit solutions
in a forthcoming paper \cite{CDWKM}.

We write the harmonic functions as a linear combination of several
harmonic functions associated with multiple centers located at $\vec
x_A$ with electric charges $q_{AI}$ and magnetic charges $p^I_A$,  
\bea
H^I(\vec x) = h^I + \sum_A \frac{p_A^I}{|\vec{x} - \vec{x}_A|} \;,\qquad
H_I(\vec x) = h_I + \sum_A \frac{q_{AI}}{|\vec{x} - \vec{x}_A|} \;, 
\eea
where the $(h^I,h_J)$ are constants and the charges are normalized according
to \eqn{eq:normcharges}. Furthermore, we recall
\bea
F_{0p}^{-I} &=& -{\rm e}^{2g} \Big[ \nabla_p Y^I + (\nabla_p g -\ft12
i {\rm e}^{2g}R(\sigma)_p) (Y^I +\bar Y^I)\Big]\,, 
\nonumber\\  
G^-_{0pI}&=& -{\rm e}^{2g} \Big[ \nabla_p F_I + (\nabla_p g -\ft12 i
{\rm e}^{2g}R(\sigma)_p )(F_I+\bar F_I)\Big]\,, 
\label{ebhd}
\eea
and hence 
\bea
F_{0p}^{-I}+F_{0p}^{+I} &=& - \nabla_p\Big[{\rm e}^{2g}  ( Y^I+
\bar Y^I) \Big]\,,  \nonumber\\
G^-_{0pI}+ G^+_{0pI}&=& 
- \nabla_p \Big[ {\rm e}^{2g}(F_I + \bar F_I)\Big] \,.
\eea
We also rewrite the expressions \eqn{dk} and \eqn{ak} in terms of the rescaled
variables,  
\bea
{\rm e}^{-\cal K} + \ft12 \chi &=& -  128 i \,{\rm e}^{3g} \,\nabla^p
\Big[ {\rm e}^{-g}\, \nabla_pg \,(F_\Upsilon-\bar F_\Upsilon)\Big] - 32 i \,{\rm e}^{6g}
\,(R(\sigma)_p)^2 (F_\Upsilon-\bar F_\Upsilon) \nonumber \\
&& -64 \,{\rm e}^{4g} R(\sigma)_p\,\nabla^p  (F_\Upsilon+\bar F_\Upsilon ) \,,
\label{dk2} \\
 H^I \stackrel{\leftrightarrow}\nabla_p H_I
&=& -\ft12 \chi  \, R(\sigma)_p \nonumber\\
&&- 128 \,  \nabla^q \Big[ 2 \nabla_{[p} g \,\nabla_{q]} (F_\Upsilon +
{\bar F}_\Upsilon)  + i 
\nabla_{[p}  \Big( {\rm e}^{2g} \, R(\sigma)_{q]}\,(F_\Upsilon - {\bar
F}_\Upsilon)   \Big) \Big] 
\,. \label{ak2}
\end{eqnarray} 
We note that both sides of \eqn{ak2} are manifestly divergence free away from
the centers. Furthermore, in the one-center case where the solution
has spherical symmetry and depends only on the radial coordinate, the terms
involving $F_\Upsilon$ and its complex conjugate vanish in
\eqn{ak2}. 

Let us first briefly discuss the solutions in the absence of 
$R^2$-interactions. Then \eqn{dk2} and \eqn{ak2} imply that 
\beq
{\rm e}^{-{\cal K}} + \ft12\chi =0\,, \qquad 
  R(\sigma)_m =  - 2 \,\chi^{-1}\, H^I
\stackrel{\leftrightarrow}\nabla_m H_I\,. 
\eeq
Once we have solved the stabilization equations, we have thus
constructed the full solution in terms of the harmonic functions. 
For the static solutions, where $R(\sigma)_m=0$, this implies that $H^I
\stackrel{\leftrightarrow}\nabla_m H_I=0$, which leads to the
following condition on the charges \cite{Sabra},
\beq
h^I \, q_{AI} - h_I \, p^I_{A} =  0 \;, \qquad
p^I_A \,q_{BI} -q_{AI} \, p^I_{B}= 0
\;. \label{charge-cond}
\eeq
The second condition implies that the charges are mutually local, i.e., 
the solution can be related to one carrying electric charges only
by electric/magnetic duality.
Moreover it implies that the
total angular momentum of a dyon $A$ in the field of a dyon $B$ vanishes.

Asymptotically, at spatial infinity, the fields can be expanded 
in powers of $1/|\vec{x}|$, 
\bea\label{eq:falloff1}
Y^I(\vec x) = Y^I(\infty) + {y^I\over |\vec{x}|} + \cdots \;, \qquad
F_I(\vec x) = F_I(\infty) + {f_I\over |\vec{x}|} + \cdots\,.
\eea
Inspection of (\ref{gen-stab-eqs}) then shows that 
$ Y^I(\infty)- \bar  Y^I(\infty) = i h^I$, $
F_I(\infty)- \bar  F_I(\infty) = i h_I $ as well as
$y^I- \bar y^I = i p^I$ and $f_I-\bar f_I =
iq_I$, where $p^I$ and $q_I$ denote the (total) magnetic 
and electric charges,
respectively. 
The homogeneity of the holomorphic function 
$F$ implies $F_I\, \delta Y^I - Y^I\,\delta F_I = 0$, and therefore 
we conclude that $y^I
F_I(\infty) - f_I Y^I(\infty) = 0$.  The following results can then be 
obtained by explicit calculation, 
\bea
\vec R(\sigma)(\vec{x}) 
&=& {\rm e}^{\cal K}\Big[h_I p^I - h^I q_I \Big]\,  {\vec{x}\over
|\vec{x}|^3} + \cdots \,, \nonumber\\
{\rm e}^{-{\cal K} -2g} &=& \Big[{\rm e}^{-{\cal K} -2g}\Big]_\infty\, 
\Big\{ 1
+ \Big[{\rm e}^{{\cal K}/2+g}\Big]_\infty \, { 2 M_{\rm ADM} \over |\vec{x}|} 
+ \cdots\Big\}\,,
\eea
where $M_{\rm ADM}$ denotes the ADM mass (in Planck units),
\begin{equation}
  \label{eq:adm_mass}
  M_{\rm ADM} = \ft12  \Big[{\rm e}^{{\cal K}/2+g}\Big]_\infty 
\left( p^I F_I(\infty) -q_I Y^I(\infty) + {\rm h.c.}\right)
  \,.
\label{adm}
\end{equation}
Note that the $M_{\rm ADM}$ can be written as $M_{\rm ADM} = \ft12 [{\bar h} 
Z(\infty) + h {\bar Z} (\infty) ]$, where $Z$ was defined in 
(\ref{holo-bps-mass}).  For static solutions ${\bar h}Z$ is real by virtue of 
the first condition in (\ref{charge-cond}), so that
$M_{\rm ADM} = {\bar h} Z(\infty) $ \cite{Sabra}.
With these results one easily shows
that the  electric and magnetic fields (\ref{ebhd}) have the 
characteristic $1/r^2$ fall-off at spatial infinity. 

We now discuss the solutions with $R^2$-interactions.
In the presence of  these interactions 
the equations 
\eqn{dk2} and \eqn{ak2} are more
difficult to analyze. We note that, generically, multi-centered
solutions satisfying  $H^I \stackrel{\leftrightarrow}\nabla_p H_I =0$
are not static, since \eqn{ak2} then reads  
\begin{eqnarray}
R(\sigma)_p = - 
256 \, \chi^{-1} \,   \nabla^q \Big[ 2 \nabla_{[p} g \,\nabla_{q]}
(F_{\Upsilon} + {\bar F}_{\Upsilon})  + i 
\nabla_{[p}  \Big( {\rm e}^{2g} \, R(\sigma)_{q]}\,(F_{\Upsilon} 
- {\bar F}_{\Upsilon})  
\Big) \Big] 
\,.
\label{rext}
\eea 
Examples of black holes 
exhibiting this feature will be discussed in \cite{CDWKM}.

When a solution has a horizon with full supersymmetry, we can connect the
results of this section to those of section
\ref{sec:N=2}.  In doing so, it is important to keep in mind that 
we used a different parametrization of the metric
in section \ref{sec:N=2}
(c.f. \ref{confm}).  The results can be connected
through the following identifications, which are valid at
the horizon (which we take to be located at $|\vec{x}|=0$, 
for convenience),
\bea
Y^I \approx
{[{\rm e}^{{\cal K}/2} \bar Z X^I]_{\rm hor}\over |\vec{x}|} \;,\qquad 
F_I(Y) \approx {[{\rm e}^{{\cal K}/2} \bar Z F_I(X)]_{\rm hor} 
\over |\vec{x}|} \,,
\nonumber\\
{\rm e}^{-g}  \approx  
{[{\rm e}^{{\cal K}/2}\vert Z \vert]_{\rm hor} \over |\vec{x}|} \;,\qquad
h  \approx
{ Z \over \vert Z\vert}\Big|_{\rm hor} \;,\qquad
\Upsilon \approx 
- \frac{64}{ |\vec{x}|^2}  \;. 
\label{yz}
\eea
In particular, when approaching
the horizon, the expressions for the field strengths \eqn{ebhd} coincide with \eqn{eq:fs4}. 

In the presence of $R^2$-interactions, the
homogeneity of the holomorphic function $F(Y,\Upsilon)$ implies
$F_I\, \delta Y^I - Y^I\, \delta F_I = 2 \Upsilon \,\delta F_{\Upsilon}$.
If we assume that, at spatial infinity, the fields $Y^I, F_I$ and 
${\rm e}^{-g}$
have an asymptotic
expansion of the type (\ref{eq:falloff1}), and if we furthermore  
assume that $\Upsilon\, \delta F_{\Upsilon}$ falls off to zero sufficiently
rapidly 
so that we have $y^I F_I(\infty) - f_I Y^I(\infty) = 0$, then the ADM mass
of the solution is still given by (\ref{adm}).

Finally, let us point out that the  multi-centered solutions we have 
studied 
can now be used as a starting point for computing the
metric on the moduli space of four-dimensional extremal black holes
in the presence of $R^2$-interactions.  In the absence of $R^2$-interactions,
it was found \cite{MSS,GP} that the metric on the moduli space
of electrically charged BPS black holes is determined in terms of a moduli
potential $\mu$ given by $\mu = - \ft{1}{2} \chi
\int d^3 \vec{x} \, {\rm e}^{-4g}$.
As shown above, 
when turning on $R^2$-interactions, ${\rm e}^{-2g}$ itself gets modified 
according to (\ref{eq:line}) with
${\rm e}^{\cal K} $ given by \eqn{dk2}.
Thus, the moduli potential receives $R^2$-corrections 
which are encoded
in ${\rm e}^{-4g}$ (and possibly further 
corrections due to additional explicit
modifications of $\mu$).  
Using \eqn{dk2} we can rewrite $\mu$ as follows,
\bea
\mu &=& - \ft{1}{2} \chi 
\int d^3 x \; {\rm e}^{-4g} \nonumber\\
&=&  
\int d^3 x \; {\rm e}^{-4g} \,
\Big( {\rm e}^{- \cal K} + 256 \,   {\rm e}^{3g} \,
\nabla_m \Big[ {\rm Im}\, F_{\Upsilon} \, \nabla^m {\rm e}^{-g}  
\Big] 
\nonumber\\
&& \qquad \qquad \qquad -\, 64 \, {\rm e}^{6g} \, (R(\sigma)_m)^2
\,{\rm Im} \,F_\Upsilon + 128\, {\rm e}^{4g} \, R(\sigma)^m \,
\nabla_m\, {\rm Re} \,F_\Upsilon \Big)  \nonumber\\ 
&=&   \int d^3 x \; {\rm e}^{-2g} \, 
\Big( i \Big[{\bar Y}^I 
F_I (Y, \Upsilon)
- {\bar F}_I ({\bar Y}, {\bar \Upsilon}) {Y}^I\Big] + 
4\,\vert\Upsilon\vert\, {\rm Im}\, F_{\Upsilon}  \Big) \, ,
\label{modulipot}
\eea
where we integrated by parts. Observe that the combination 
$i \Big[{\bar Y}^I 
F_I (Y, \Upsilon)
- {\bar F}_I ({\bar Y}, {\bar \Upsilon}) { Y}^I\Big] + 
4\,\vert\Upsilon\vert\, {\rm Im}\, F_{\Upsilon}$, 
when evaluated at the horizon of a
BPS black hole, is precisely equal to $\pi^{-1} r^{-2}$ times 
the expression for its
macroscopic entropy (see (\ref{entropy}) and (\ref{yz}))! 
This intriguing feature of  (\ref{modulipot}) may indicate that there are no
additional explicit modifications of $\mu$ due to $R^2$-interactions.
This is currently under investigation \cite{CDWKM}. In establishing
(\ref{modulipot}) we dropped 
certain boundary terms when integrating by parts. Some of those are
known to be proportional to $\vert\vec{x}_A - \vec{x}_B\vert^{-1}$
(for two non-coincident  centers $A$ and $B$) and therefore do not
contribute to the metric on the moduli space \cite{MSS}.  

\acknowledgments

During the course of this work various institutes contributed and
offered us hospitality. 
B.d.W. thanks the Aspen Center for Physics in the context
of attending the workshop ``String Dualities and Their 
Applications''. J.K. is financially 
supported by the FOM foundation. T.M. thanks 
the Erwin Schr\"odinger International Institute for
Mathematical Physics in the context of attending the workshop
``Duality, String Theory and M-Theory''. 
This work was supported in part by the European Commission TMR
programme ERBFMRX-CT96-0045.


\appendix

\setcounter{equation}{0}
\section{Notation and conventions}\label{a}

We denote spacetime indices by
$\mu ,\nu , \cdots,$ and Lorentz indices by 
$a,b, \cdots = 0,1,2,3$.
Indices $i,j,k, \ldots$ are usually reserved for SU(2)-indices. 
Our (anti)symmetrization conventions are
\beq
[ab] = \ft12 ( ab-ba )\;\;,\;\; (ab) = \ft{1}{2} ( ab + ba ) \;\;\;.
\eeq
We take
\beq
\g_a \g_b =\eta_{ab} + \g_{ab}\,, \qquad 
\g_5 = i \g_0\g_1\g_2\g_3\,,
\eeq
where $\eta_{ab}$ is of signature $(-+++)$. The complete antisymmetric
tensor satisfies
\beq
\ve^{abcd} = e^{-1} \ve^{\mu\nu\l\s} e_\mu^a e_\nu^b e_\l^c 
e_\s^d\,,   \qquad
\ve^{0123} = i \,,
\eeq
which implies 
\beq
\g_{ab} = -\ft12\ve_{abcd}\g^{cd}\g_5\,.
\eeq
The dual of an antisymmetric tensor field $F_{ab}$ is given by
\beq
\tilde{F}_{ab} = \ft12\ve_{abcd}F^{cd}       \,,
\eeq
and the (anti)selfdual part of $F_{ab}$ reads
\beq
 F^{\pm}_{ab} = \ft12(F_{ab}\pm\tilde{F}_{ab})  \,.
\eeq
We note the following useful 
identities for (anti)selfdual tensors in 4 dimensions:
\beqa
 G^\pm_{[a[c}\, H_{d]b]}^\pm &=& \pm\ft18 G^\pm_{ef} \,H^{\pm ef} 
\,\varepsilon_{abcd} -\ft14( G^\pm_{ab}\, H^\pm_{cd} +G^\pm_{cd}\, 
H^\pm_{ab}) \,,\nonumber \\ 
G^{\pm}_{ab} \, H^{\mp cd} + G^{\pm cd} \, H^{\mp}_{ab} &=& 4 
\d^{[c}_{[a} G^{\pm}_{b]e} \, H^{\mp d]e} \,,\nonumber \\
\ft12 \varepsilon^{abcd} \,G^{\pm}_{[c}{}^e \, H^\pm_{d]e} &=& \pm  
G^{\pm [a}{}_{e} \, H^{\pm b]e}\,,\nonumber \\ 
G^{\pm ac}\,H^\pm_c{}^b + G^{\pm bc}\,H^\pm_c{}^a &=& -\ft12 \eta^{ab}\, 
G^{\pm cd}\,H^\pm_{cd} \,,\nonumber \\ 
G^{\pm ac}\,H^\mp_c{}^b &=&G^{\pm bc}\,H^\mp_c{}^a\,, \qquad 
G^{\pm ab}\,H^\mp_{ab} =0 \,. 
\eeqa
Note that under hermitian conjugation (${\rm h.c.}$) selfdual becomes 
antiselfdual
and vice-versa. Any SU(2) index $i$ or any quaternionic
index $\a$ changes position under ${\rm h.c.}$, for instance
\beq (T_{ab\, ij})^* = T_{ab}^{ij} \,,
\qquad \qquad (A_i^\a)^* = A^i_\a\,.
\eeq  

\section{Superconformal calculus}\label{b}
\setcounter{equation}{0}

The superconformal algebra consists of general coordinate, local
Lorentz, dilatation, special conformal, chiral U(1) and SU(2), and
$Q$- and $S$-supersymmetry transformations. The fully supercovariant
derivatives are denoted by $D_a$. 
We use ${\cal D}_\mu$ to denote a covariant derivative with respect to
Lorentz, dilatation, chiral U(1), SU(2) and gauge transformations. 
The component fields of the various superconformal multiplets carry
certain Weyl 
and chiral weights.  Those of the Weyl multiplet and of the 
supersymmetry transformation parameters are listed in table 
\ref{weyl}, whereas those
of the vector and of hypermultiplets are given in table \ref{vechyp}.
These tables also list the fermion chirality of the various  
component fields.
To exhibit the form of the derivatives ${\cal D}_\m$ and the
normalization of the gauge fields contained in them,  let us give the
derivative of the chiral spinor $\e^i$,
\beq
{\cal D}_\m \e^i = \pa_\m \e^i - \ft14 \omega_\m^{ab} \,\gamma_{ab} \e^i +
\ft12 (b_\m + i A_\m) \e^i + \ft 12 {\cal V}_\m{}^{\!i}{}_{\!j} \e^j \,.
\eeq

\TABLE[t]{
\begin{tabular}{|c||cccccccc|ccc||cc|}
\hline
&
   \multicolumn{11}{c||}{Weyl multiplet} &
   \multicolumn{2}{c|}{parameters} \\
\hline
\hline
field          &
   $e_\mu{}^a$   &
   $\psi_\mu^i$  &
   $b_\mu$       &
   $A_\mu$       &
   ${{\cal V}_\mu}^i{}_j$ &
   $T_{ab}^{ij}$    &
   $\chi^i$      &
   $D$           &
   $\omega_\mu^{ab}$ &
   $f_\mu{}^a$   &
   $\phi_\mu^i$  &
   $\e^i$        &
   $\eta^i$ \\[.5mm]
\hline
\hline     
$w$         & $-1$     & $-\ft12$  & $0$      & $0$      & $0$
  & $1$      &  $\ft32$  & $2$      & $0$       & $1$      &
$\ft12$  & $-\ft12$ & $\ft12$ \\[.5mm]
 \hline
$c$         & $0$      & $-\ft12$  & $0$      & $0$      & $0$
  & $-1$     &  $-\ft12$ & $0$      & $0$       & $0$      &
$-\ft12$ & $-\ft12$ & $-\ft12$ \\[.5mm]
\hline
$\gamma_5$ & & $+$ &&&&& $+$ &&&& $-$ & $+$ & $-$ \\[.5mm]
\hline
\end{tabular}\\[.13in]
\caption{Weyl and chiral weights
($w$ and $c$, respectively)
             and fermion chirality ($\gamma_5$)
             of the Weyl multiplet component fields
             and of the supersymmetry transformation parameters.}
\label{weyl}
}

The gauge fields for Lorentz and $S$-supersymmetry transformations
are composite objects and given by
\beqa \omega_\mu^{ab} &=& -2e^{\nu[a}\partial_{[\mu}e_{\nu]}{}^{b]}
     -e^{\nu[a}e^{b]\s}e_{\mu c}\partial_\s e_\nu{}^c
     -2e_\mu{}^{[a}e^{b]\nu}b_\nu   \nonumber\\
     & & -\ft{1}{4}(2\bar{\psi}_\mu^i\g^{[a}\psi_i^{b]}
     +\bar{\psi}^{ai}\g_\mu\psi^b_i+{\rm h.c.}) \,,\nonumber\\
   \phi_\mu^i &=& \ft12(\g^{\rho\s}\g_\mu-\ft{1}{3}\g_\mu\g^{\rho\s}) 
     \Big({\cal D}_\rho\psi_\s^i-\ft{1}{16}T^{ab\,ij}\g_{ab}\g_\rho\psi_{\s j}
     +\ft{1}{4}\g_{\rho\s}\chi^i\Big) \nonumber \\
&=& -\ft13( 4\d_\m^{[\rho}\g^{\sigma]} + 
\varepsilon_{\m\l}{}^{\rho\sigma}\g^\l ) 
     \Big({\cal D}_\rho\psi_\s^i-\ft{1}{16}T^{ab\,ij}\g_{ab}\g_\rho\psi_{\s j}
     +\ft{1}{4}\g_{\rho\s}\chi^i\Big) \,,
\label{dependent}
\eeqa
respectively.  
The gauge field for special conformal transformations
is also a composite object and was already given in \eqn{f-def}, up to
fermionic terms.
There is no need to give the transformation rules for the dependent
gauge fields. The explicit transformation rule for $\phi_\m^i$ is,
however, used in the calculations of this paper and we have 
presented it in \eqn{phi-transfo}. 
Throughout this paper we need certain supercovariant curvature tensors,
\beqa R(Q)_{\mu\nu}^i &=&
     2{\cal D}_{[\mu}\psi_{\nu]}^i
     -\g_{[\mu}\phi_{\nu]}^i
     -\ft{1}{8} T^{ab\,ij}\g_{ab} \g_{[\mu}\psi_{\nu]j}\,, \nonumber\\
     R(A)_{\mu\nu} &=&
     2\partial_{[\mu}A_{\nu]}
     -i\Big(\ft{1}{2}\bar{\psi}_{[\mu}^i\phi_{\nu]i}
     +\ft{3}{4}\bar{\psi}_{[\mu}^i\g_{\nu]}\chi_i-{\rm 
 h.c.}\Big)\,, \nonumber\\
     R({\cal V})_{\mu\nu}\,^i{}_j &=& 2\partial_{[\mu} {\cal V}_{\nu ]}^i{}_j 
     + {\cal V}_{[\mu }^i{}_k{\cal V}_{\nu ]}^k{}_j \nonumber\\
     && +\Big(2\bar{\psi}_{[\mu }^i\phi_{\nu ]j} 
     -3\bar{\psi}_{[\mu }^i\g_{\nu]}\chi_j
     -({\rm h.c.}\,;\,{\rm traceless}) \Big)\,, \nonumber \\ 
     R(M)_{\mu\nu}^{ab}&=& R(\omega)_{\m\n}^{ab} - 4 f_{[\m}^{\,[a}
     e_{\n]}^{\, b]} + \Big(\ft12 \bar\psi^i_{[\m} \g^{ab}
     \phi_{\n]i}  + {\rm h.c.}   \Big) \nonumber\\ && + 
   \Big(\ft12 \bar\psi^i_{[\m} T_{ij}^{ab} \phi_{\n]}^i-\ft34
     \bar\psi^i_{[\m}\g_{\n]} \g^{ab} \chi^i - \bar\psi^i_{[\m}\g_{\n]}
     R_{\mu\nu}(Q)_i + {\rm h.c.}   \Big) \,, \\
     R(S)_{\mu\nu}^i & = &  2{\cal D}_{[\mu}\phi_{\nu]}^i  
       -2 f_{[\mu}^a\g_a\psi_{\nu]}^i
     -\ft18 D \llap/ T_{ab}^{ij}\g^{ab}\g_{[\mu} \psi_{\nu]\, j}
     -3 \bar{\chi}_j\g^a\psi_{[\mu}^j\g_a\psi_{\nu]}^i \nonumber\\
     & & +\ft14 R({\cal V})_{ab}{}^i_{\ j}\g^{ab}\g_{[\mu}\psi_{\nu]}^j
     +\ft12 i R(A)_{ab}\g^{ab}\g_{[\mu}\psi_{\nu]}^i  \,, \nonumber\\
      R(D)_{\mu\nu} & = & 2 \pa_{[\mu} b_{\nu]} -2
      f_{[\mu}^a\,e_{\nu] a}  -\Big(\ft12  
      \bar{\psi}_{[\mu}^i\phi_{\nu] i}
      +\ft34\bar{\psi}^i_{[\mu}\g_{\nu]}\chi_i
        +{\rm h.c.} \Big)
      \,.\nonumber
\eeqa

The following modified curvature tensors appear in the component
fields of the chiral multiplet $W^2$ (c.f. \ref{background-def}), 
\beqa
{\cal R}(M)_{ab}{}^{\!cd} & = &    R(M)_{ab}\,^{cd} + \ft1{16}\Big( 
T^{ijcd}\,T_{ijab} + T^{ij}_{ab}\, T^{cd}_{ij}  \Big)\,, \nonumber \\
 {\cal R}(S)^i_{ab} & = & R(S)^i_{ab} + \ft34
 T_{ab}^{ij} \chi_j\,.
\eeqa
The $T^2$-modification cancels exactly the $T^2$-terms in the 
contribution to ${\cal R}(M)$ from $f^{\,a}_\m$. 
The curvature ${\cal R}(M)_{ab}{}^{\!cd}$ satisfies the following
relations, 
\beqa
{\cal R}(M)_{\m\n}{}^{\!ab} \,e^\n{}_b &=& i \tilde R(A)_\m{}^a +\ft32
D\,e_\m{}^a\,, \nonumber\\
\ft14 \varepsilon_{ab}{}^{ef} \,\varepsilon^{cd}{}_{gh} \,{\cal 
R}(M)_{ef}{}^{\!gh} &=& {\cal R}(M)_{ab}{}^{\!cd} \,,\nonumber\\
\varepsilon_{cdea}\,{\cal R}(M)^{cd\,e}{}_{\!b} &=& 
\varepsilon_{becd} \,{\cal R}(M)_a{}^{\!e\,cd}  = 2\tilde{R}_{ab}(D) =
2i R_{ab}(A)\,. 
\label{RM-constraints}
\eeqa
The first one is the constraint that determines the field $f_\m{}^a$
while the remaining equations are Bianchi identities. Note that the
modified curvature does not satisfy the pair exchange property, 
\bea
{\cal R}(M)_{ab}{}^{\!cd} = {\cal R}(M)^{cd}{}_{\!ab} + 4 i
\d^{[c}_{[a} \, \tilde R(A) _{b]}^{~}{}^{\!d]}\,.
\eea
{From} these
equations one determines for instance
\beq
{\cal R}(M)^\pm_{0[p}{}^0{}_{q]}^{~} = \pm\ft12 i R(A)^\pm_{pq}\,.
\eeq
We note that  $R(Q)_{ab}^i$ satisfies the constraint 
\bea
\g^\m R(Q)_{\m\n}^i + \ft32 \g_\n \chi^i =0\,,
\label{constrq}
\eea
which must therefore hold for its variation as well. This constraint
implies that $R(Q)_{\m\n}^i$ is anti-selfdual, as follows from
contracting it with $\g^\n\,\g_{ab}$.

The curvature  ${\cal R}(S)^i_{ab}$  satisfies
\beq
\g^a\tilde {\cal R}(S)^i_{ab} = 2\, D^a\tilde R(Q)_{ab}^i\,,
\eeq
as a result of the Bianchi identities and of the constraint (\ref{constrq}).
This identity (upon contraction with $\g^b\g_{cd}$) leads to 
\beq
{\cal R}(S)_{ab}^i-\tilde{\cal R}(S)_{ab}^i = 2\,\rlap/\!D(
R(Q)^i_{ab} + \ft34 \g_{ab} \chi^i)\,.
\eeq
%
%

\TABLE[t]{
\hspace{0.5cm}
\begin{tabular}{|c||cccc||cc|}
\hline
&
   \multicolumn{4}{c||}{vector multiplet}  &
   \multicolumn{2}{l|}{hypermultiplet } \\  
\hline
\hline
field          &
   $X^I$         &
   $\O_i^{\,I}$      &
   $W_\mu^{\,I}$     &
   $Y_{ij}^{\,I}$    &
   $A_i^\a$    &
   $\zeta^\a$  \\
\hline
\hline 
$w$         & $1$ & $\ft32$  & $0$ & $2$ &
            $1$ & $\ft32$  \\[.5mm]
\hline
c         & $-1$& $-\ft12$ & $0$ & $0$ &
            $0$ & $-\ft12$ \\[.5mm]
\hline
$\gamma_5$&     &  $+$     &     &     &
                &  $-$     \\[.5mm]
\hline
\end{tabular}
\caption{Weyl and chiral weights ($w$ and $c$,
  respectively) and fermion chirality ($\gamma_5$) of the vector and
  hypermultiplet component fields.}
\label{vechyp}
\hspace{0.5cm}
}

%
\end{document}